\documentclass[10pt,a4paper,journal,english]{IEEEtran}

\usepackage[T1]{fontenc}
\usepackage[utf8]{inputenc}

\usepackage{babel}
\usepackage[subrefformat=parens,skip=6pt]{subcaption}
\usepackage{graphicx}
\usepackage{rotating}
\usepackage{multirow}
\usepackage{amscd}
\usepackage[centertags,reqno]{amsmath}
\usepackage{amssymb}
\usepackage{theorem}
\usepackage{paralist}
\usepackage{verbatim}
\usepackage{balance}
\usepackage{cite}
\usepackage[multiple]{footmisc}

\allowdisplaybreaks

\newcommand{\ceil}[1]{\ensuremath{\left\lceil#1\right\rceil}}

\newcommand{\prob}[1]{\ensuremath{\Pr\left\{#1\right\}}}
\renewcommand{\vec}[1]{\ensuremath{\mathbf{#1}}}
\newcommand{\sw}[2]{\ensuremath{_{#1}^{#2}}}
\newcommand{\field}[1]{\ensuremath{\mathbb{#1}}}

\newcommand{\sequence}[3]{\ensuremath{(#1_{#2}, \ldots, #1_{#3-1})}}
\newcommand{\seq}[2]{\ensuremath{\vec{#1}\sw{0}{#2}}}

\newcommand{\Pri}{\ensuremath{P_\mathrm{i}}}
\newcommand{\Prd}{\ensuremath{P_\mathrm{d}}}
\newcommand{\Prs}{\ensuremath{P_\mathrm{s}}}
\newcommand{\Prt}{\ensuremath{P_\mathrm{t}}}
\newcommand{\Prr}{\ensuremath{P_\mathrm{r}}}

\newlength\figwidth
\newcommand{\insertfig}[3]{%
   \begin{figure}[!t]
      \centering
      \includegraphics[width=\figwidth]{#2}
      \caption{#3}
      \label{#1}
   \end{figure}}

\newcommand{\inserttab}[4]{%
   \begin{table}[!t]
      \begin{minipage}{\columnwidth}
         \caption{#3}
         \label{#1}
         \centering
         \renewcommand{\arraystretch}{1.15}
         \begin{tabular}{#2}
            \hline
            #4
            \hline
         \end{tabular}
      \end{minipage}
   \end{table}}

\newcommand{\inserttabd}[4]{%
   \begin{table*}[!t]
      \begin{minipage}{\textwidth}
         \caption{#3}
         \label{#1}
         \centering
         \renewcommand{\arraystretch}{1.15}
         \begin{tabular}{#2}
            \hline
            #4
            \hline
         \end{tabular}
      \end{minipage}
   \end{table*}}

\title{{GPU} Implementation and Optimization
   of a Flexible {MAP} Decoder
   for Synchronization Correction}

\author{%
   Johann~A.~Briffa
   \thanks{%
      Manuscript submitted February 13, 2014;
      accepted May 5, 2014 for publication in
      the IET Journal of Engineering.
      Parts of this research have been carried out using computational
      facilities procured through the European Regional Development Fund,
      Project ERDF-080,
      and a gift from NVIDIA.
      }%
   \thanks{%
      J.~A.~Briffa is with the
      Dept.\ of Computing, University of Surrey, Guildford GU2 7XH, England.
      Email: j.briffa@surrey.ac.uk
      }%
   }

\begin{document}

\maketitle

\begin{abstract}
In this paper we present an optimized parallel implementation of a flexible
MAP decoder for synchronization error correcting codes, supporting a very
wide range of code sizes and channel conditions.
On mid-range GPUs we demonstrate decoding speedups of more than two orders
of magnitude over a CPU implementation of the same optimized algorithm,
and more than an order of magnitude over our earlier GPU implementation.
The prominent challenge is to maintain high parallelization efficiency over
a wide range of code sizes and channel conditions, and different execution
hardware.
We ensure this with a dynamic strategy for choosing parallel execution
parameters at run-time.
We also present a variant that trades off some decoding speed for significantly
reduced memory requirement, with no loss to the decoder's error correction
performance.
The increased throughput of our implementation and its ability to work with
less memory allow us to analyse larger codes and poorer channel conditions,
and makes practical use of such codes more feasible.
\end{abstract}



\section{Introduction}
\label{sec:introduction}


The problem of correcting synchronization errors has recently seen an increase
in interest \cite{mbt2010commsurveys}.
We believe this is due to two factors: recent applications for such codes,
where traditional techniques for synchronization cannot be applied, and
the feasibility of decoding due to improvements in computing resources.
A recent application is for bit-patterned media \cite{hdek2010transcomm,
hdke2007transmag}, where written-in errors can be modeled as synchronization
errors.
Bit-patterned media is of great interest to the magnetic recording industry
due to the potential increase in writing density.
Another example is robust digital watermarking, where a message is embedded
into a media file and an attacker seeks to make the message unreadable.
An effective attack is to cause loss of synchronization;
synchronization-correcting codes have been successfully applied to resist these
attacks in speech \cite{cs08tifs} and image \cite{bbds11iscas} watermarking.

Most practical decoders for synchronization correction work by extending the
state space of the underlying code to account for the state of the channel
(which represents the synchronization error).
This increases the decoding complexity significantly, particularly under
poor channel conditions where the state space is necessarily larger.
While optimal decoding is achievable, the complexity involved remains a
barrier for wider adoption.
The problem is even more pronounced when these codes are part of an iteratively
decoded construction.


A key practical synchronization-correcting scheme is the concatenated
construction by Davey and MacKay \cite{dm01ids}, where the inner code tracks
synchronization on an unbounded random insertion and deletion channel.
We presented a maximum a-posteriori (MAP) decoder for a generalized
construction of the inner code in \cite{bsw10icc}, and improved encodings
in \cite{bb11isit}.
In \cite{briffa13jcomml} we presented a parallel implementation of our MAP
decoder on a graphics processing unit (GPU) using NVIDIA's Compute Unified
Device Architecture (CUDA) \cite{cuda-pg-50}.
This resulted in a decoding speedup of up to two orders of magnitude,
depending on code parameters and channel conditions.
Since that work we have also presented a number of additional improvements
to the MAP decoder algorithm \cite{bbw14joe}, resulting in a speedup of
over an order of magnitude in a serial implementation, as we shall show.
Unfortunately, these algorithmic improvements change the proportion of time
spent computing the various equations, so that a straightforward application
of the algorithm improvements to our earlier GPU implementation does not
yield the expected speedup.
A more careful parallelization strategy is required, which we discuss in
this paper.


Other GPU implementations of the forward-backward algorithm (on which our
MAP decoder is based) have significant fundamental differences from this work.
Recent work on turbo decoders \cite{lwk2010cases, wsw2011jsps, xcjgb2013iwcmc}
is limited to the parallel implementation of specific binary code
constructions.
Further, turbo decoders work with a small state-space of a fixed size, and
the branch metric is trivial to compute, so is generally recomputed as needed.
These factors simplify the problem considerably, making it much easier to
optimize a parallel implementation.
In contrast, we consider the problem of efficiently parallelizing a flexible
MAP decoder implementation, where:
\begin{inparaenum}[\em a)]
\item the code size is variable,
\item codes are non-binary, and the implementation works with variable
alphabet size,
\item the state space is variable in size, depending on channel conditions,
and can easily run into hundreds of states for poor channels, and
\item the branch metric computation requires a lattice traversal, and
represents a significant fraction of the overall complexity.
\end{inparaenum}
Due to these variables, hard optimizations cannot be done, so our challenge
is to plan things to work well with suitable run-time decisions.
Since various optimization decisions are taken at run-time, this also allows
us to automatically cater for different hardware.


In this paper we present an optimized parallel implementation of the MAP
decoder with the algorithmic improvements of \cite{bbw14joe}.
Two variants of this algorithm are implemented: a direct implementation where
intermediate metrics are stored for the decoding of a whole frame, and a
reduced-memory implementation where some intermediate metrics are recalculated
for the backward pass.
This considerably reduces the memory footprint of the decoder, which is a
particularly important consideration for a GPU implementation, at the cost of
decoding speed.
In optimizing the implementation, we consider the use of GPU on-chip memory
to reduce memory transfers and to improve access patterns.
We also introduce a dynamic strategy for choosing how each function is
parallelized, including the number of threads to use, in order to optimize
the efficiency and usage of the GPU compute cores.


The approach presented here can be generalized to other implementations of
the forward-backward algorithm, such as those used in turbo decoding and in
applications of hidden Markov models.
The techniques we propose here are particularly useful in similarly flexible
implementations, commonly found in simulators and other research tools.
In addition, we discuss a number of techniques for handling device
limitations and optimizing the parallelization parameters dynamically;
these may be relevant to other researchers working with CUDA.


This paper starts with definitions and a summary of earlier work in
Section~\ref{sec:preliminaries}, including descriptions of the coding scheme,
channel, and the MAP decoder algorithm.
Our contributions start in Section~\ref{sec:challenges} where we consider
the challenges for an efficient parallel implementation of the MAP decoder.
This is followed with an initial parallel implementation and a reduced-memory
variant in Section~\ref{sec:parallelization}.
In Section~\ref{sec:advanced} we ensure that the parallel decoder works
within hardware constraints and improve efficiency when code parameters are
less than ideal.
We analyze the practical performance on three GPU systems
in Section~\ref{sec:analysis}, followed by final results in
Section~\ref{sec:results} and conclusions in Section~\ref{sec:closure}.


\section{Preliminaries}
\label{sec:preliminaries}


\subsection{The Coding Scheme}

In this paper we are concerned with the coding scheme of \cite{briffa13jcomml,
bbw14joe}, which we summarize below.
The encoding is defined by the sequence $\mathcal{C}=\sequence{C}{0}{N}$,
which consists of the constituent encodings
$C_i: \mathbb{F}_q \hookrightarrow \mathbb{F}_2^n$ for $i=0,\ldots,N-1$,
where $n,q,N \in \mathbb{N}$, $2^n \geq q$, and $\hookrightarrow$ denotes
an injective mapping.
For any sequence $\vec{z}$, denote arbitrary subsequences as
$\vec{z}\sw{a}{b} = \sequence{z}{a}{b}$, where $\vec{z}\sw{a}{a} = ()$ is an
empty sequence.
Given a message $\vec{D}\sw{0}{N} = \sequence{D}{0}{N}$, each $C_i$ maps
the $q$-ary message symbol $D_i \in \mathbb{F}_q$ to codeword $C_i(D_i)$
of length $n$.
That is, $\vec{D}\sw{0}{N}$ is encoded as
$\vec{X}\sw{0}{nN} = C_0(D_0) \| \cdots \| C_{N-1}(D_{N-1})$,
where $\vec{y}\|\vec{z}$ is the juxtaposition of $\vec{y}$ and $\vec{z}$.

The above encoding is normally used as an inner code to correct synchronization
errors, serially concatenated with a conventional outer code to correct
residual substitution errors.
In such a construction, the MAP decoder's posterior probabilities are used
to initialize the outer decoder.
Such a construction can be iteratively decoded by setting the prior symbol
probabilities of the MAP decoder with extrinsic information from the previous
pass of the outer decoder.


\subsection{Channel Model}

As in \cite{briffa13jcomml, bbw14joe} we consider the Binary Substitution,
Insertion, and Deletion (BSID) channel, an abstract random channel with
unbounded synchronization and substitution errors.
This channel was originally presented in \cite{bahl75} and more recently used
in \cite{dm01ids, ratzer05telecom, bs08isita, bsw10icc, bb11isit} and others.
At \emph{time} $t$, one bit enters the channel, and one of three events
may happen:
insertion with probability $\Pri$ where a random bit is output;
deletion with probability $\Prd$ where the input is discarded; or
transmission with probability $\Prt=1-\Pri-\Prd$.
A substitution occurs in a transmitted bit with probability $\Prs$.
After an insertion the channel remains at time $t$ and is subject to
the same events again, otherwise it proceeds to time $t+1$, ready for
another input bit.

We define the \emph{drift} $S_t$ at time $t$ as the difference between
the number of received bits and the number of transmitted bits before the
events of time $t$ are considered.
As in \cite{dm01ids}, the channel can be seen as a Markov process with the
state being the drift $S_t$.
It is helpful to see the sequence of states as a trellis diagram, observing
that there may be more than one way to achieve each state transition.


\subsection{The MAP Decoder}
\label{sec:decoder}

We summarize here the optimized MAP decoder of \cite{bbw14joe}, which we are
concerned with parallelizing.
The decoder uses the standard forward-backward algorithm for hidden Markov
models.
We assume a message sequence $\vec{D}\sw{0}{N}$, encoded to the sequence
$\seq{X}{\tau}$, where $\tau = nN$.
The sequence $\seq{X}{\tau}$ is transmitted over the BSID channel, resulting
in the received sequence $\seq{Y}{\rho}$, where in general $\rho$ is not
equal to $\tau$.
To avoid ambiguity, we refer to the message sequence as a \emph{message}
of size $N$ and the encoded sequence as a \emph{frame} of size $\tau$.
We calculate the APP $L_i(D)$ of having encoded symbol $D \in \field{F}_q$ in
position $i$ for $0 \leq i < N$, given the entire received sequence, using
\begin{align}
   \label{eqn:L}
   L_i(D) & = \frac{1}{\lambda_N(\rho-\tau)} \sum_{m',m} \sigma_i(m',m,D)
   \text{,}\\
   \text{where~}
   \label{eqn:lambda}
   \lambda_i(m) &= \alpha_i(m) \beta_i(m)
   \text{,}\\
   \label{eqn:sigma}
   \sigma_i(m',m,D) &= \alpha_i(m') \gamma_i(m',m,D) \beta_{i+1}(m)
   \text{,}
\end{align}
and $\alpha_i(m)$, $\beta_i(m)$, and $\gamma_i(m',m,D)$ are the forward,
backward, and state transition metrics respectively.
Note that strictly, the above metrics depend on $\seq{Y}{\rho}$, but for
brevity we do not indicate this dependence in the notation.
The summation in \eqref{eqn:L} is taken over the combination of $m',m$,
being respectively the drift before and after the symbol at index $i$.
The forward and backward metrics are obtained recursively using
\begin{align}
   \label{eqn:alpha}
   \alpha_i(m) &= \sum_{m',D} \alpha_{i-1}(m') \gamma_{i-1}(m',m,D)\text{,}\\
   \label{eqn:beta}
   \text{and~}
   \beta_i(m) &= \sum_{m',D} \beta_{i+1}(m') \gamma_i(m,m',D)\text{.}
\end{align}
Initial conditions are given by $\alpha_0(m)$ and $\beta_N(m)$, set as the
prior probabilities of the frame boundaries.
Finally, the state transition metric is defined as
\begin{equation}
\label{eqn:gamma}
\gamma_i(m',m,D) =
   \prob{ D_i = D }
   R( \vec{Y}\sw{ni+m'}{n(i+1)+m} \mid C_i(D) )
\end{equation}
where $C_i(D)$ is the $n$-bit sequence encoding $D$ and
$R( \vec{\dot{y}} | \vec{x} )$ is the probability of receiving a sequence
$\vec{\dot{y}}$ given that $\vec{x}$ was sent through the channel (we refer
to this as the receiver metric).
The \emph{a priori} probability $\prob{ D_i=D }$ is determined by the source
statistics, which we generally assume to be equiprobable so that
$\prob{ D_i=D } = 1/q$.
In iterative decoding, the prior probabilities are set using extrinsic
information from the previous pass of an outer decoder.

Since the set of all possible states is unbounded for the channel considered,
a practical implementation has to take sums over a finite subset, chosen so
that only the least likely states are omitted.
For a transmitted segment of length $T$ bits we denote the range of states
considered by the upper and lower limits $m_T^{+}$, $m_T^{-}$ respectively,
for a state space of size $M_T = m_T^{+} - m_T^{-} + 1$.
Therefore, the $\alpha$ and $\beta$ metrics are defined over a state space
of size $M_\tau = m_\tau^{+} - m_\tau^{-} + 1$ while the $\gamma$ metric is
defined over a state space of size $M_\tau$ for the initial drift $m'$ and a
state space of size $M_n = m_n^{+} - m_n^{-} + 1$ for the drift change $m-m'$.
The precise determination of the size of the state space is considered
in \cite{bbw14joe}.


The receiver metric $R( \vec{\dot{y}} | \vec{x} )$ is computed using a
recursion over a lattice as follows.
The required lattice has $n+1$ rows and $\dot\mu+1$ columns, where $\dot{\mu}$
is the length of $\vec{\dot{y}}$, and $n$ is the length of $\vec{x}$.
Each horizontal path represents an insertion with probability
$\frac{1}{2}\Pri$, each vertical path is a deletion with probability $\Prd$,
while each diagonal path is a transmission with probability $\Prt\Prs$ if
the corresponding elements from $\vec{x}$ and $\vec{\dot{y}}$ are different
or $\Prt(1-\Prs)$ if they are the same.
Let $F_{i,j}$ represent the lattice node in row $i$, column $j$.
Then the lattice computation in the general case is defined by the recursion
\begin{equation}
\label{eqn:F}
F_{i,j} =
   \frac{1}{2}\Pri F_{i,j-1} +
   \Prd F_{i-1,j} +
   \dot{Q}(\dot{y}_j | x_i) F_{i-1,j-1}
   \text{,}
\end{equation}
which is valid for $i<n$, and where $\dot{Q}(y|x)$ can be directly computed
from $y,x$ and the channel parameters:
\begin{equation}
\begin{split}
   &\dot{Q}( y | x) =
   \begin{cases}
      \Prt\Prs      & \text{if $y \neq x$} \\
      \Prt(1-\Prs)  & \text{if $y = x$}\text{.} \\
   \end{cases}
\end{split}
\end{equation}
Initial conditions are given by
\begin{equation}
\begin{split}
   &F_{i,j} =
   \begin{cases}
      1 & \text{if $i=0$, $j=0$} \\
      0 & \text{if $i<0$ or $j<0$.} \\
   \end{cases}
\end{split}
\end{equation}
The last row is computed differently as the channel model does not allow
the last event to be an insertion.
In this case, when $i=n$, the lattice computation is defined by
\begin{equation}
\label{eqn:F_lastrow}
F_{n,j} =
   \Prd F_{n-1,j} +
   \dot{Q}(\dot{y}_j | x_n) F_{n-1,j-1}
   \text{.}
\end{equation}
Finally, the required receiver metric is obtained from this computation as
$R( \vec{\dot{y}} | \vec{x} ) = F_{n,\dot{\mu}}$.


Observe that for a given $\vec{x}$, the receiver metric
$R( \vec{\dot{y}} | \vec{x} )$ needs to be determined for all subsequences
$\dot{\vec{y}}$ within the drift limit considered.
Therefore, for a given symbol $D$ and initial drift $m'$ in \eqref{eqn:gamma},
the lattice computation is only done once for the largest drift change $m-m'$
that needs to be considered.
The required values for the remaining values of $m$ are then also available
in the last row of the lattice.


Note also that the horizontal distance of a lattice node from the main diagonal
is equivalent to the channel drift for the corresponding transmitted bit.
For the transmitted sequence of $n$ bits considered, we can take advantage
of this by limiting the lattice computation to paths within a fixed corridor
of width $M_n$ around the main diagonal.


Finally, note that the $\alpha$ and $\beta$ metrics are normalized as they
are computed to avoid exceeding the limits of floating-point representation.
Specifically, for $\alpha$ the computation \eqref{eqn:alpha} is changed to:
\begin{align}
\label{eqn:alpha_norm}
\alpha_i(m) &= \frac{\alpha'_i(m)}
   {\sum_{m'} \alpha'_i(m')}
   \text{,} \\
\label{eqn:alpha_prenorm}
\text{where~}
\alpha'_i(m) &= \sum_{m',D}
   \alpha_{i-1}(m') \gamma_{i-1}(m',m,D)
   \text{.}
\end{align}
A similar argument applies for the computation of $\beta$.
Also, the receiver metric is computed at single precision\footnote{%
We refer to 32-bit floating point as single precision, and 64-bit floating
point as double precision.},
while the remaining equations use double precision.


\subsection{CUDA Notation}

In this paper we follow the usual CUDA notation, which we summarize here for
convenience.
For further detail, the reader is referred to \cite{cuda-pg-50}.
CUDA defines a general-purpose parallel programming model for a hybrid system
with a \emph{host} CPU and an attached GPU \emph{device} (or more than one).
The device contains the GPU chip, organized as a number of
\emph{multiprocessors} with a fixed number of compute cores each, and
off-chip memory.
Each multiprocessor also contains a fixed amount of on-chip \emph{shared}
memory, accessible by all compute cores in the multiprocessor.
Off-chip memory is accessible by all GPU threads through \emph{global} memory
variables in read/write mode, or in read-only mode through \emph{constant}
or \emph{texture} memory constructs.

Every function executed on the GPU is called a \emph{kernel}; this is run
as a \emph{grid} of equally-shaped \emph{blocks} of parallel threads, as
specified by the execution configuration.
To avoid ambiguity, we avoid using the term \emph{block} for any other purpose.
Each block of threads executes on the same multiprocessor in groups of
threads called \emph{warps}.
Threads in a given warp start execution at the same address but are free
to branch an execute independently (i.e.\ \emph{diverge}).
However, highest efficiency is achieved when there is no divergence within
a warp.
Note that more than one block may be \emph{resident} in a given multiprocessor
if sufficient resources (i.e.\ registers and shared memory) are available.
This increases the number of warps available to the scheduler, and may be used
to hide latency.


\section{Challenges for Parallel Implementation}
\label{sec:challenges}

\subsection{Effect of Changes to Receiver Metric Computation}


As compared with our previous GPU implementation, the optimized decoder
summarized in Section~\ref{sec:decoder} changes to the way the receiver
metric is computed.
Of particular relevance to this work, in the optimized decoder
\begin{inparaenum}[\em a)]
\item for a given $\vec{x}$, we simultaneously compute
$R( \vec{\dot{y}} | \vec{x} )$ for all subsequences of $\vec{\dot{y}}$, and
\item we replace the trellis-based forward pass of \cite{bsw10icc} with a
more efficient corridor-constrained lattice implementation.
\end{inparaenum}
Together, these changes result in a decrease in complexity (for computing
the receiver metric) of up to three orders of magnitude, depending on
channel conditions.

\begin{figure}[tb]
   \centering
   \begin{subfigure}[b]{\figwidth}
      \includegraphics[width=\figwidth]{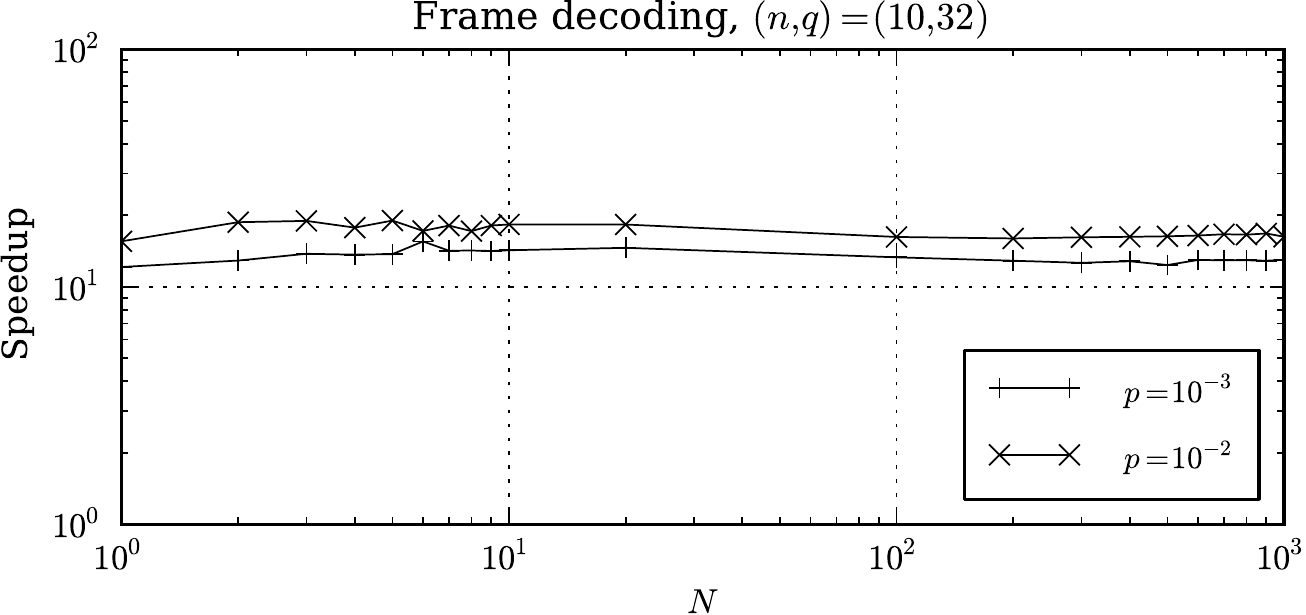}
      \caption{}
      \label{fig:speedup-oldvnew-cpu-N}
   \end{subfigure}
   \begin{subfigure}[b]{\figwidth}
      \includegraphics[width=\figwidth]{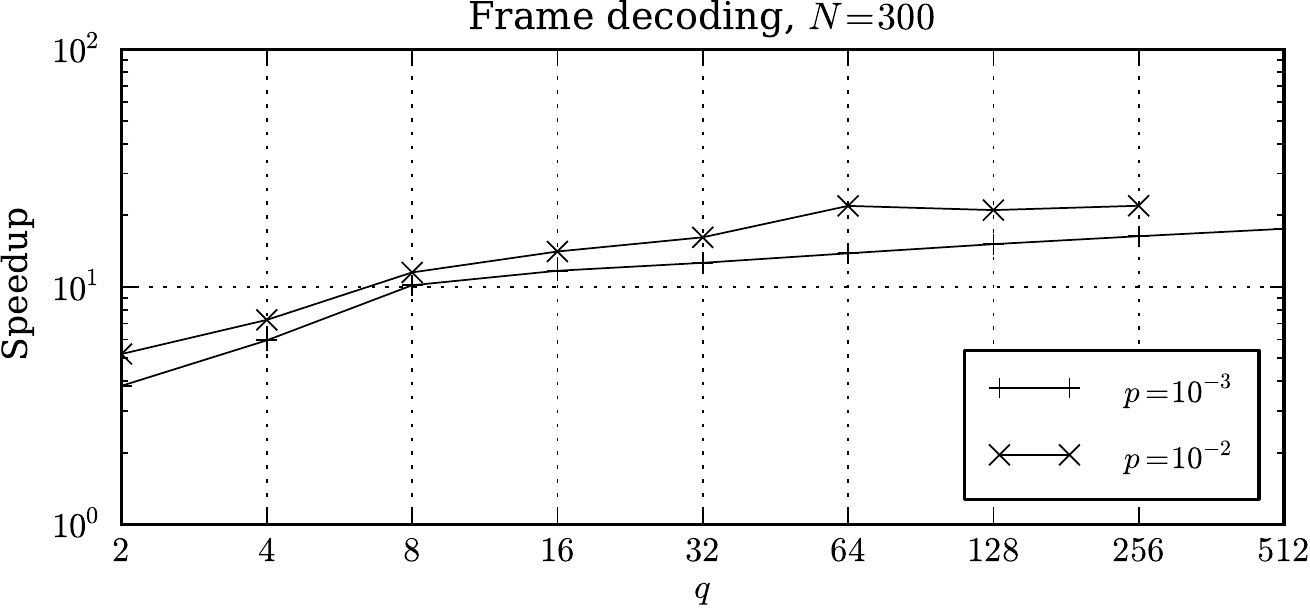}
      \caption{}
      \label{fig:speedup-oldvnew-cpu-q}
   \end{subfigure}
   \caption{
      Decoding speedup of a serial CPU implementation for lattice corridor
      batch computation (new) over individual trellis computation (old)
      of receiver metric, at different channel conditions $p:=\Pri=\Prd;
      \Prs=0$ for a range of
      \subref{fig:speedup-oldvnew-cpu-N} message size $N$ and
      \subref{fig:speedup-oldvnew-cpu-q} alphabet size $q$.
      }
   \label{fig:speedup-oldvnew-cpu}
\end{figure}

To demonstrate the effect on the overall decoding speed, we repeat the
simulations of \cite{briffa13jcomml} using a serial CPU implementation of
the improved decoder.
Results comparing the overall decoding speed, for the same codes and on the
same computer, are shown in Fig.~\ref{fig:speedup-oldvnew-cpu}.
Under moderate channel conditions it can be seen that the improved decoder is
more than ten times faster over a wide range of message size $N$ and alphabet
size $q$.
Speedup improves for poorer channel conditions.


Such a significant change means that the receiver metric computation can no
longer be considered to dominate the overall complexity.
This is particularly so for a GPU implementation, where we expect a
higher speedup in computing the $\gamma$ metric since this has greater
data parallelism.
Therefore, in parallelizing this improved algorithm we also have to pay
particular attention to an efficient parallel computation of the $\alpha$
and $\beta$ metrics, which have internal dependencies.

\subsection{Effect of Implementation Flexibility}

The GPU implementation of this MAP decoder also faces challenges not considered
in similar parallel implementations for turbo codes \cite{lwk2010cases,
wsw2011jsps, xcjgb2013iwcmc}.
In our case, the flexibility of the MAP decoder and the nature of the channel
we are concerned with pose additional difficulties for parallelization,
as noted in Section~\ref{sec:introduction}.
The reference CPU implementation is intended for simulation applications, and
consequently supports a wide range of code parameters.
Specifically, the CPU implementation supports:
\begin{inparaenum}[\em a)]
\item arbitrary codeword length $n \geq 1$,
\item channel symbols from an arbitrary finite alphabet, as long as the
alphabet is defined in an additive Abelian group,
\item messages defined over an arbitrary finite alphabet of size $q \geq 2$,
as long as each message symbol is representable by a codeword,
i.e.\ $q \leq s^n$ for a channel symbol alphabet of size $s$,
\item arbitrary message length $N \geq 1$,
\item arbitrary state space limits $m_n^{-} \leq 0$, $m_n^{+} \geq 0$,
$m_\tau^{-} \leq m_n^{-}$, $m_\tau^{+} \geq m_n^{+}$, chosen dynamically based
on channel conditions,
\item arbitrary received sequence length $\rho$, as long as the drift at the
end falls within the chosen state space limits,
i.e.\ $m_\tau^{-} \leq \rho-\tau \leq m_\tau^{+}$,
\item arbitrary specification of constituent encodings $C_i$,
\item arbitrary thresholds to avoid pursuing low-probability trellis paths
(not considered in this study), and
\item lazy computation of the $\gamma$ metric (not considered in this study;
potentially useful with path thresholding).
\end{inparaenum}

Additionally, the implementation supports the independent choice of real
number types for the computation of the forward-backward algorithm and the
inner lattice traversal.
On the CPU implementation, supported types include single and double
precision IEEE floating point numbers, as well as GNU multi-precision numbers
\cite{gnump5}.
This allows the user to trade off arithmetic speed and storage requirements
for accuracy, and the comparison of results with an exact implementation.
The implementation also allows the user to choose between different algorithms
to compute the receiver metric.
These include the constrained-corridor lattice algorithm of
Section~\ref{sec:decoder}, an unconstrained version of the algorithm,
and the trellis-based algorithm used in our earlier work.

All these variables pose a number of particular challenges for an efficient
parallel implementation.
First of all, it is important for the parallel implementation to support the
same range of parameters as the CPU implementation.
This can be a problem when functions are parallelized across one of these
variables, as the required range may exceed constraints of the parallel
architecture.
For example, in our earlier GPU implementation, $q$ was limited by hardware
constraints, and results were given for alphabet sizes up to $q=512$.
Another expected problem is that the parallelization efficiency depends on the
code parameters.
With so many variables and such a wide range for each, it becomes particularly
problematic to ensure high parallelization efficiency under all conditions.
Unlike our earlier GPU implementation, in this work we propose specific
steps to improve efficiency under suboptimal conditions.
This variability also has an effect on scheduling kernel issues across
multiple streams, as the time taken by individual kernels depends on the
code parameters.
On most current hardware, the ideal issue order for different kernels that can
execute concurrently depends on their relative timings \cite{rennich2011gtc}.
When such decisions had to be taken in this work we have favoured larger
codes, where the speedup is of greater objective benefit.

A more subtle problem is that functions which have many distinct execution
paths tend to require a higher register count.
In a parallel implementation, this can reduce the upper limit for parallel
threads, as each multiprocessor will have a finite register bank to share.
Further, the compiler may reduce register usage by `spilling' some values
to global memory, for a very significant increase in latency.
This problem can be alleviated in part by using C++ templates \cite{eckel2000,
eckel2003}.
Each template instance is effectively independent, so that the corresponding
execution path decisions are taken at compile time.
This reduces register requirements, which is of great benefit in the parallel
implementation; while also useful for the CPU implementation, the difference
in execution speed is minimal.
However, this solution is not without its cost: the use of templates increases
the compiler's work in proportion to the number of combinations of template
parameters.
This can increase object code size and compilation time considerably.
Further, for a CUDA compilation with current tools, this also presents
a constraint due to the limit on constant memory use within the same
compilation unit.
The solution we have applied to this problem was to divide the instantiation
of the various combinations of template parameters into separate units.
That is, the class methods were specified in a separate implementation header,
imported by a number of source files, each of which instantiated an independent
subset of all required template parameter combinations.

Finally, it is worth realizing that the requirements of a CPU implementation
can be at odds with those of a GPU implementation.
For example, the sequence of constituent encodings $\mathcal{C}$ is held
in memory as a vector of $N$ constituent encodings $C_i$, each of which is
a vector of $q$ codewords, where each codeword is in turn a vector of $n$
channel symbols.
This is ideal for a CPU implementation, allowing each symbol to be accessed
by indirect addressing, and also allowing references to individual codewords
or codebooks to be specified naturally.
On the GPU, however, such an organization is problematic because it requires
$Nq$ independent memory allocations and memory transfers, each of which is
an expensive operation.
It is more efficient to reorganize the table on the CPU, copying the data
into a single flat array and using that structure on the GPU.
This avoids changing the canonical data structure on the CPU, which would
have wider repercussions, and is the solution we have adopted in this work.


\section{Initial Parallel Implementation}
\label{sec:parallelization}


\subsection{Global Storage}
\label{sec:global_storage}

The most straighforward implementation of the MAP decoder considers it
as consisting of four functions, one for computing each of the $\gamma$,
$\alpha$, $\beta$, and $L$ metrics.
These depend on each other, dictating the order of computation: specifically,
$\alpha$ and $\beta$ depend on $\gamma$, while $L$ depends on $\alpha$,
$\beta$ and $\gamma$.
Therefore, $\gamma$ needs to be computed first, followed by $\alpha$ and
$\beta$ (in any order, as these are independent of each other), and finally
$L$ is computed.
This methodology, which we refer to as \emph{global} storage, assumes that
there is sufficient memory to store the $\alpha$, $\beta$ and $\gamma$ metrics.

\subsubsection{$\gamma$ metric}
\label{sec:gamma_global}

Starting with the computation of the $\gamma$ metric \eqref{eqn:gamma},
we have already observed in \cite{briffa13jcomml} that the computation is
independent for each of $i$, $D$, $m'$, and $m$.
Furthermore, as noted in Section~\ref{sec:decoder}, for any given $i$, $D$,
and $m'$, we only need to perform the lattice computation for the largest
drift change $m-m'$ to be considered.
It makes sense, therefore, to compute and store the $\gamma$ metric for the
whole valid range of $m$, for any given $i$, $D$, and $m'$.


To make the best use of the available data parallelism, we initially use
block coordinates $(i,m')$ for grid size $N \times M_\tau$, and threads with
coordinate $D$ for block size $q$.
This increases the parallelism by a factor $M_\tau$ with respect to our earlier
GPU implementation, where the $\gamma$ computation kernel had a grid size $N$.
This increased parallelism is particularly useful when global storage is
not possible, as we shall see in Section~\ref{sec:local_storage}.
Each thread performs an independent lattice computation and determines the
$\gamma$ metric for the whole valid range of $m$ (i.e.\ $M_n$ entries).


As in our earlier GPU implementation, we store the four-dimensional $\gamma$
matrix as a flat array in global memory.
However, we change the indexing order, so as to have index $D$ innermost.
This allows the threads in a warp to access a contiguous range of memory;
this access is fully coalesced as long as the initial address and the
transaction size are a multiple of 128 bytes.
Since $q$ is always a power of two, this is guaranteed for any $q \geq 16$
with double precision storage.
This effect is particularly valuable since there is no warp divergence, as the
lattice traversal is identical for all $D$.
This index is followed by $m-m'$, so that consecutive accesses by the same
thread are as close to each other as possible, maximizing cache re-use for
small $q$.


In order to minimize global memory access and avoid register spilling into
local memory, each thread holds the current lattice row being computed in
shared memory.
This requires an array of size $n + m_n^{+} + 1$ single precision numbers
per thread, dynamically allocated on kernel launch.
The lattice computation algorithm is re-written accordingly.

\subsubsection{$\alpha$ metric}
\label{sec:alpha_global}

The $\alpha$ metric computation can be divided into two main operations:
the main computation \eqref{eqn:alpha_prenorm}, followed by normalization
\eqref{eqn:alpha_norm}.
We have already observed in \cite{briffa13jcomml} that the main computation
at index $i$ and state $m$ depends on normalization at index $i-1$ for all
$m'$, while normalization at index $i$ for state $m$ depends on computation
at index $i$ for all $m'$.
Also, for any given $i$, computation and normalization are independent for
different values of $m$.


Parallelization strategy is the same as our earlier GPU implementation, where
a separate kernel call is required for the main computation at each $i$,
followed by a separate kernel to perform normalization at that $i$.
This is necessary because the only way to synchronize across a grid is the
completion of a kernel call \cite{cuda-pg-50}.

The main computation at $i$ uses block coordinate $m$ for a grid size of
$M_\tau$ and threads with coordinate $D$ for block size $q$, where each
thread computes the corresponding partial summation over $m'$.
The final result is then computed from these partial sums using a parallel
summation across the threads in the block, using shared memory to communicate
between threads.
This requires a shared memory array of $q$ double precision values per block.
Normalization requires two steps: computing the sum of all $\alpha'_i$,
and dividing each $\alpha'_i$ by this sum.
Both are most easily parallelized across a single block of $M_\tau$ threads.
This uses only one multiprocessor but greatly facilitates implementation
in a function which corresponds to a very small proportion of the overall
computation time.

In contrast with our earlier GPU implementation, we extract the initialization
of the $\alpha$ metric at $i=0$ to a separate kernel.
Since initialization consists simply of setting each of $M_\tau$ values, this
is implemented as a single block of $M_\tau$ threads.
While this change may not seem very significant, it slightly simplifies the
main computation, keeping register usage to a minimum.
In turn, this allows us to maximize occupancy by allowing more resident kernels.


The $\alpha$ metric is stored as a two-dimensional array in global memory,
with the state index $m$ innermost.
This speeds up access in the alpha metric computation kernel, where each thread
needs to read the metric values at all states for the previous index $i-1$.
%
%
The speedup is achieved by copying the row at index $i-1$ to shared memory,
requiring a shared memory array of $M_\tau$ double precision values per block.
This memory copy can be done in parallel across the block, so that global
memory access is coalesced.
The use of a two-dimensional array ensures this by correctly aligning each row.

\subsubsection{$\beta$ metric}
\label{sec:beta_global}

A similar argument applies to the computation of $\beta$, except that now
the main computation at index $i$ depends on normalization at index $i+1$.
%
%
Further, $\alpha$ and $\beta$ can be computed concurrently as there is no
data dependency between them; this can be achieved using streams on devices
that support concurrent kernel execution.

Recall that the computation of $\alpha$ and $\beta$ requires a number of
consecutive kernel calls each.
Specifically, for $\alpha$, this sequence consists of the intialization
kernel, normalization at $i=0$, computation at $i=1$, normalization at $i=1$,
and repeating computation and normalization for increasing $i$ until $i=N$.
Since each kernel depends on the completion of the preceding one, this
dependency is best expressed by issuing the kernels in the same stream.
For $\beta$, the sequence is the same but the index order is reversed
(i.e.\ initialization and normalization for $i=N$, followed by computation
and normalization pairs for decreasing $i$ from $i=N-1$ to $i=0$).
The independence of the $\beta$ kernels from the $\alpha$ kernels is expressed
by issuing these in a second stream.

Unfortunately, hardware limitations in Fermi and initial Kepler
devices\footnote{GK104 architecture, for compute capability 3.0.} cause
additional complication.
Since these devices have only one compute engine queue, if any stream has
more than one kernel scheduled consecutively, the issuer will stall until
the last kernel in the sequence is dispatched \cite{rennich2011gtc}.
Since the kernels for $\alpha$ and $\beta$ at each index have the same
complexity, we avoid this problem by using a breadth-first launch order,
as follows.
Issue first the initialization of $\alpha_{i=0}$ in stream one and of
$\beta_{i=N}$ in stream two, followed by the normalization of $\alpha_{i=0}$
in stream one and of $\beta_{i=N}$ in stream two.
This is followed by the computation of $\alpha_{i=1}$ in stream one and of
$\beta_{i=N-1}$ in stream two, and the normalization of $\alpha_{i=1}$
in stream one and of $\beta_{i=N-1}$ in stream two.
This is repeated, incrementing $i$ for $\alpha$ and decrementing for $\beta$.
Concurrent execution improves device utilization when the grid size for a
single kernel call is small.
Note that this problem does not exist in the latest Kepler architecture
(GK110), which has 32 compute engine queues \cite{nvidia-kepler}.

\subsubsection{$L$ metric}

Finally, the $L$ computation \eqref{eqn:L} is independent across $i$, $D$.
%
%
As in our earlier GPU implementation, we parallelize this across blocks with
index $i$ for a grid size $N$ and threads with index $D$ for block size $q$.
%
%
In this work, however, we also avoid multiple global memory reads and ensure
coalesced memory access by first copying the required rows at $\alpha_i$
and $\beta_{i+1}$ to shared memory.
This requires two shared memory arrays of $M_\tau$ double precision values
per block.
Each $\gamma$ value is only read once, and this is done in an order that
ensures coalesced access.

\subsection{Local Storage}
\label{sec:local_storage}


Unfortunately, \emph{global} storage is only possible when there is sufficient
memory to store the $\alpha$, $\beta$ and $\gamma$ metrics.
The required storage capacity increases with increasing $N$, $n$, $q$,
and poorer channel conditions (since this increases the required state space).
To illustrate this, consider a system with a moderate message size $N=210$
and a range of alphabet sizes $q$; we plot the required metric storage memory
in Fig.~\ref{fig:requirements-memory} over a range of channel conditions.
\insertfig{fig:requirements-memory}{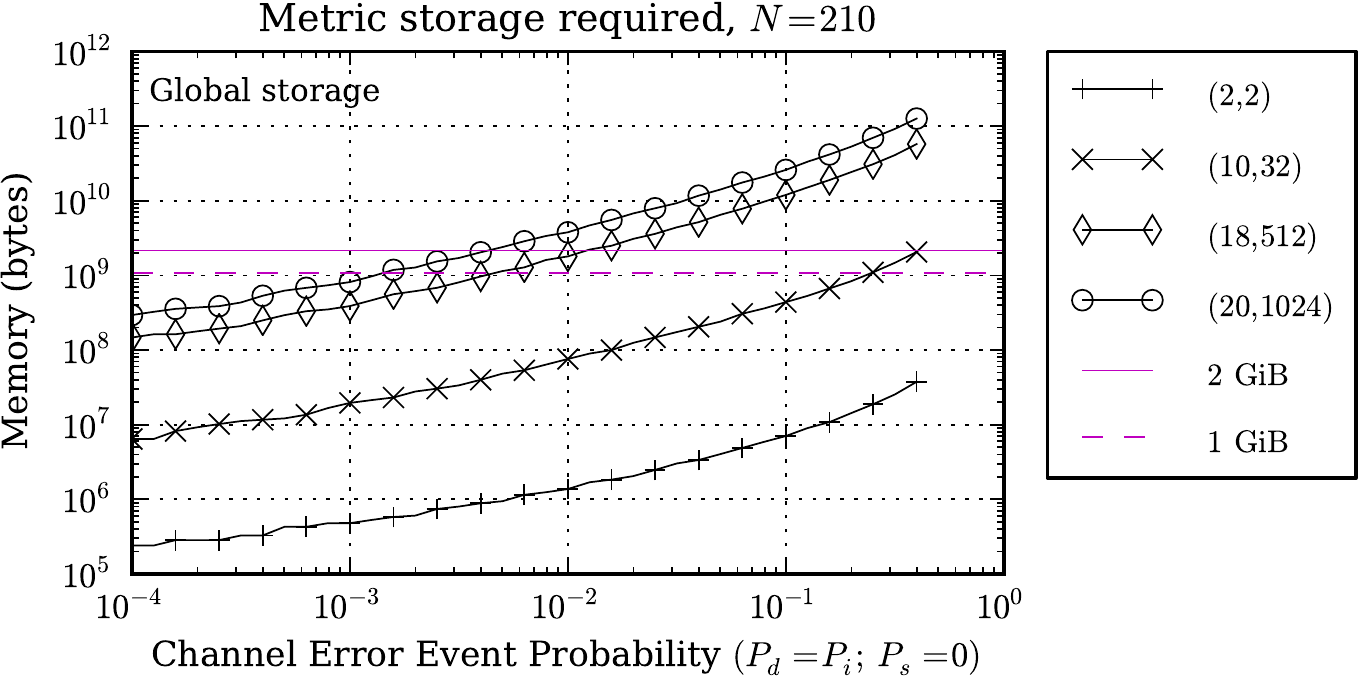}
   {Metric storage requirements for a moderate message size $N=210$ and
   half-rate codewords $(n,q)$, over a range of channel conditions.
   It is assumed that all metrics are stored as double precision values.
   }
For ease of comparison, horizontal lines are included at values of
$1\,\mathrm{GiB}$ and $2\,\mathrm{GiB}$, corresponding to common per-CPU
core or per-GPU device limits for metric storage.
It can be readily seen that these limits are reached at moderate to low
channel error rates for larger alphabet sizes, and also at high channel
error rates for moderate alphabet sizes.

\begin{figure}[tb]
   \centering
   \begin{subfigure}[b]{\figwidth}
      \includegraphics[width=\figwidth]{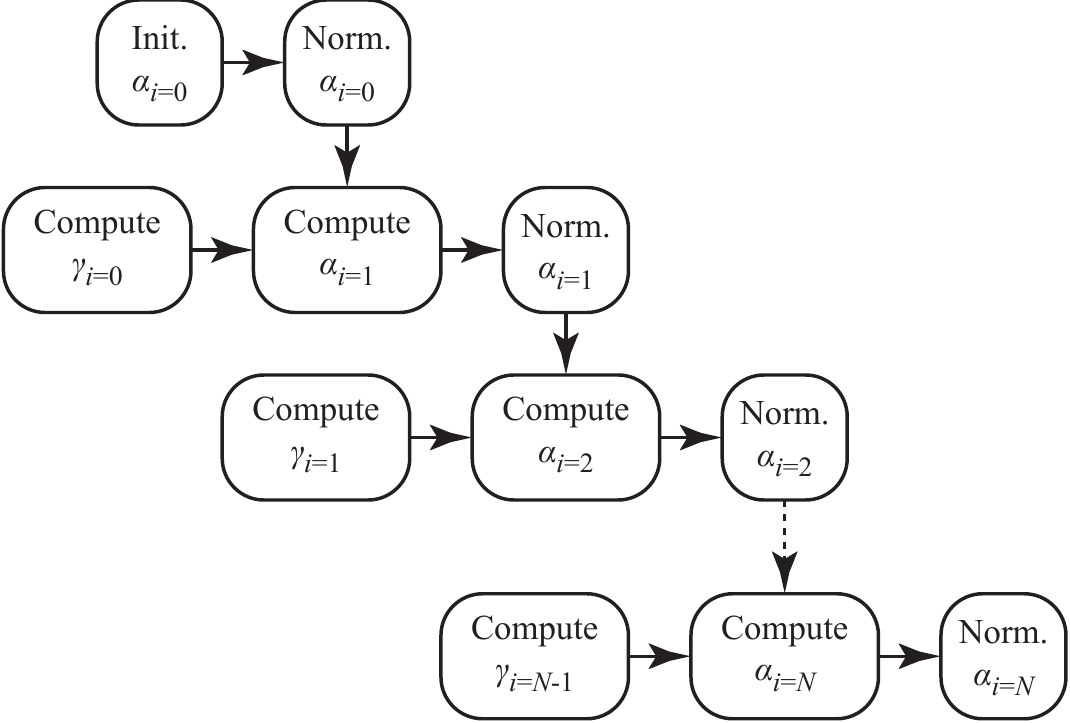}
      \caption{}
      \label{fig:dependencies-alpha}
   \end{subfigure}
   \begin{subfigure}[b]{\figwidth}
      \includegraphics[width=\figwidth]{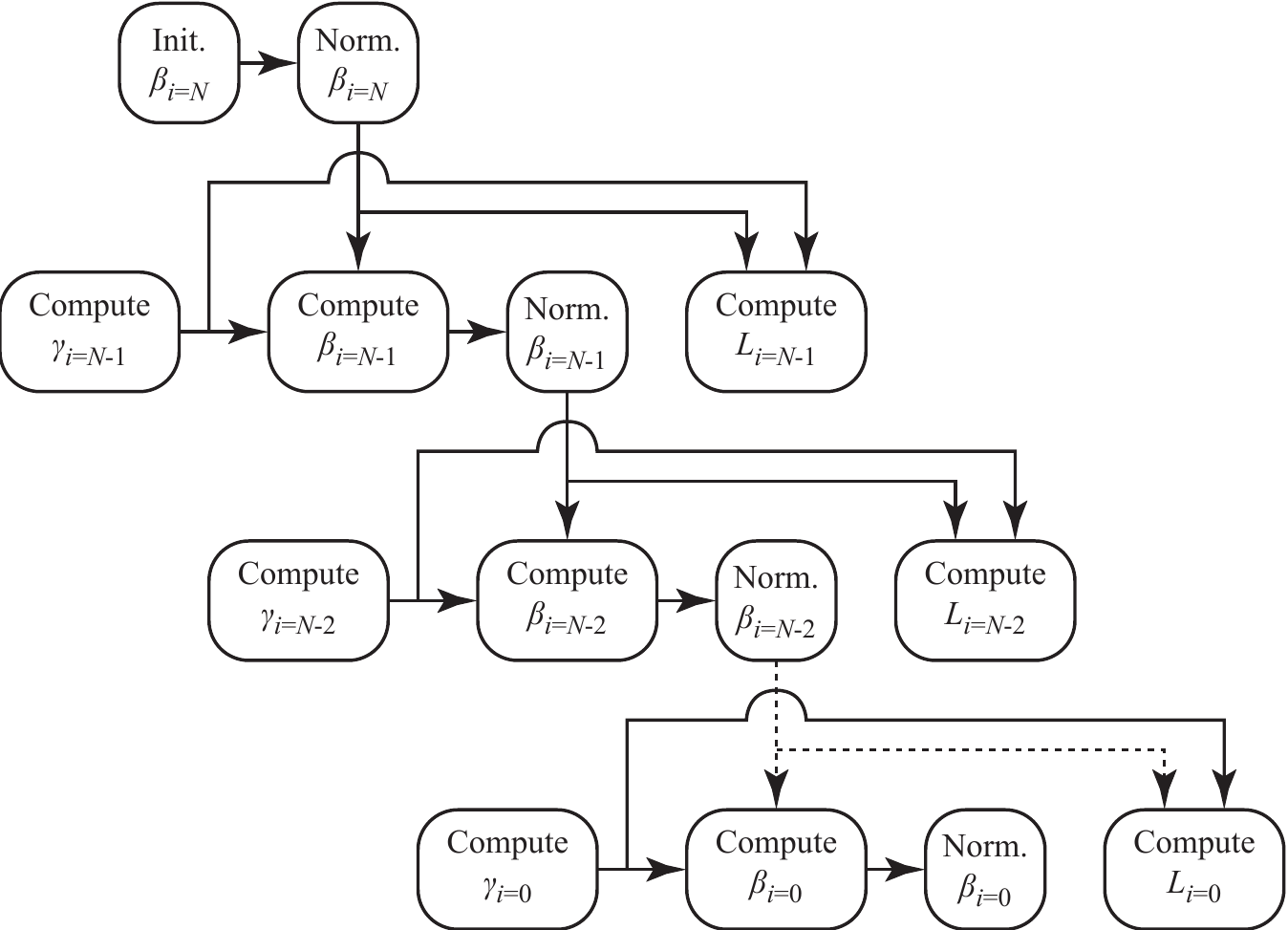}
      \caption{}
      \label{fig:dependencies-beta}
   \end{subfigure}
   \caption{
      Graphical representation of the interdependencies and the sequence of
      kernel calls required to compute
      \subref{fig:dependencies-alpha} the $\alpha$ metric, and
      \subref{fig:dependencies-beta} the $\beta$ and $L$ metrics
      in local storage mode.
      }
   \label{fig:dependencies}
\end{figure}

This problem can be resolved by dividing the computation of $\gamma_i$ across
$i$ and observing that each of $\alpha_i$, $\beta_i$ and $L_i$ depend only
on a single index for $\gamma$.
Specifically,
\begin{inparaenum}[\em a)]
\item $\alpha_i$ depends on $\alpha_{i-1}$ and $\gamma_{i-1}$,
\item $\beta_i$ depends on $\beta_{i+1}$ and $\gamma_i$, and
\item $L_i$ depends on $\alpha_i$, $\beta_{i+1}$ and $\gamma_i$.
\end{inparaenum}
Since the order of computation of $\alpha_i$ and $\beta_i$ is enforced by
their internal dependencies, each $\gamma_i$ has to be computed at least twice.
We avoid computing it a third time by first completing the $\alpha$ metric
computation, then combining the computation of $L$ with that of $\beta$.
This is shown graphically in Fig.~\ref{fig:dependencies}.


This mode requires us to divide the $\gamma$ and $L$ kernels to compute values
at a single index $i$.
Therefore the $\gamma$ compute kernel now uses block coordinate $m'$ for
grid size $M_\tau$.
Similarly, the $L$ compute kernel now only uses a single block.
In both cases, $i$ is passed as a kernel parameter.
This significantly reduces device utilization, particularly under good channel
conditions where $M_\tau$ is relatively small.


We can mitigate this problem using concurrent execution by issuing multiple
kernels, as follows.
Consider first the computation of the $\alpha$ metric, requiring kernel calls
as shown in Fig.~\ref{fig:dependencies-alpha}.
In this figure, each row corresponds to the computation of a particular row
of the $\alpha$ metric, with index $i$.
Each kernel within a given row depends on the previous kernel, while the
compute kernel also depends on the normalization kernel of the previous row.
The former dependency can be expressed by issuing the three kernels of a
given row in the same stream; the latter dependency can be expressed using
CUDA events \cite{rennich2011gtc}.
Such an arrangement allows compute kernels for the $\gamma$ metric to execute
concurrently with other kernels from previous rows, increasing device usage.

In this way, kernel issues could be divided over $N+1$ different streams.
Observe, however, that independent storage space is required for the $\gamma$
metric at each index that may be computed independently.
This makes it impractical to use a separate stream for each possible index,
in which case the amount of space required is the same as for global mode.
Instead, we limit the number of streams to a depth of four, used on a rotating
basis.
This limit was chosen empirically; others have already observed that
it is difficult to have more than four kernels running concurrently
\cite{rennich2011gtc}.

The use of streams raises the question of how best to order kernel issues,
as considered for the concurrent computation of $\alpha$ and $\beta$ in
Section~\ref{sec:global_storage}.
In this case, the problem is somewhat more complex, as different streams
may be executing kernels of different size and duration.
Additionally, the use of events to synchronize between streams requires that
the event on stream $i$ is recorded before the wait on that event is issued
on stream $i+1$.
Now for the event to be recorded, all prior kernels on that stream need to
be issued (though not necessarily executed).
Effectively, this means that for stream $i$ we need to issue the following
sequence consecutively:
   wait for event completion on stream $i-1$ $\rightarrow$
   compute $\alpha_i$ $\rightarrow$
   normalize $\alpha_i$ $\rightarrow$
   record event on stream $i$.
All this has to be issued before we can repeat the issue sequence for the
next stream.
This requirement forbids us from issuing all kernels across the four streams
in breadth-first order.
Instead, we issue the $\gamma$ computation kernels in breadth-first, then
issue the above sequences for each stream in the order required.
This maximizes concurrency of execution for the $\gamma$ computation
kernels, but limits concurrency of the remaining kernels to the combination
of computation and normalization of $\alpha$ (this applies to Fermi and
first-generation Kepler devices).

A similar argument applies for the computation of the $\beta$ and $L$ metrics,
requiring kernel calls as shown in Fig.~\ref{fig:dependencies-beta}.
Other than the order of index traversal, the only difference in this case is
the presence of the computation kernel for the $L$ metric.
The simplest way to deal with this is to issue this kernel after the
normalization kernel in the same row.
This automatically satisfies its dependencies, which are the same as for the
compute kernel for the $\beta$ metric in the same row.
It also allows the $\beta$ computation and normalization to occur first,
minimizing delays to subsequent rows.
The $L$ computation kernel has a significant duration but only has one block;
this inefficiency is easily hidden by concurrent kernels from subsequent rows.
In this case, we again issue the $\gamma$ computation kernels breadth-first,
then issue the remaining sequence for each stream in the order required.
We include the $L$ computation in this sequence rather than issuing it
separately in breadth-first order.
This allows that kernel to be concurrent with the $\beta$ kernels rather than
only with itself, making more efficient use of the available multiprocessors.


A summary of the kernels required for the initial parallel implementation
described so far, for global and local storage, is given in
Table~\ref{tab:kernels-initial}.
\inserttab{tab:kernels-initial}{lcccc}
   {Summary of the kernels required for the initial parallel implementation,
   for global and local storage.}{
   \emph{Kernel}
      & \emph{Storage Mode}
      & \emph{Grid size}
      & \emph{Block size}
      & \emph{Calls} \\
   \hline
   \multirow{2}{*}{Compute $\gamma$}
      & global
      & $N \times M_\tau$
      & $q$
      & $1$
      \\
      & local
      & $1 \times M_\tau$
      & $q$
      & $N$
      \\
   \hline
   Initialize $\alpha$,$\beta$
      & both
      & $1$
      & $M_\tau$
      & $1$
      \\
   Compute $\alpha$,$\beta$
      & both
      & $M_\tau$
      & $q$
      & $N$
      \\
   Normalize $\alpha$,$\beta$
      & both
      & $1$
      & $M_\tau$
      & $N+1$
      \\
   \hline
   \multirow{2}{*}{Compute $L$}
      & global
      & $N$
      & $q$
      & $1$
      \\
      & local
      & $1$
      & $q$
      & $N$
      \\
   \hline
   Compute $\Phi_T$
      & both
      & $1$
      & $M_\tau$
      & $2$
      \\
   }
It can be seen that the block size is equal to $q$ or $M_\tau$.


\section{Advanced Considerations}
\label{sec:advanced}

\subsection{Handling Resource Limits}
\label{sec:aggregation}


The initial parallel implementation described previously assumes that
it is possible to issue kernels with the given block sizes: $q$ for the
$\gamma$, $\alpha$, $\beta$, and $L$ computation kernels, and $M_\tau$
for the remaining kernels.
However, this may not be possible due to a number of constraints:
\begin{inparaenum}[\em a)]
\item the device limit on the maximum number of threads per block,
\item register pressure, which depends on both the number of registers needed
per thread and the number of registers available per multiprocessor, and
\item shared memory pressure, which depends on the requirements for a given
block size and the available shared memory per multiprocessor.
\end{inparaenum}
Limits for a device depend on its compute capability, and can be queried by the
implementation at run-time \cite{cuda-pg-50}.

To illustrate the shared memory requirements, consider again a system with
$N=210$ and a range of alphabet sizes $q$; we plot the required shared
memory per block in Fig.~\ref{fig:requirements-shared} over a range of
channel conditions.
\insertfig{fig:requirements-shared}{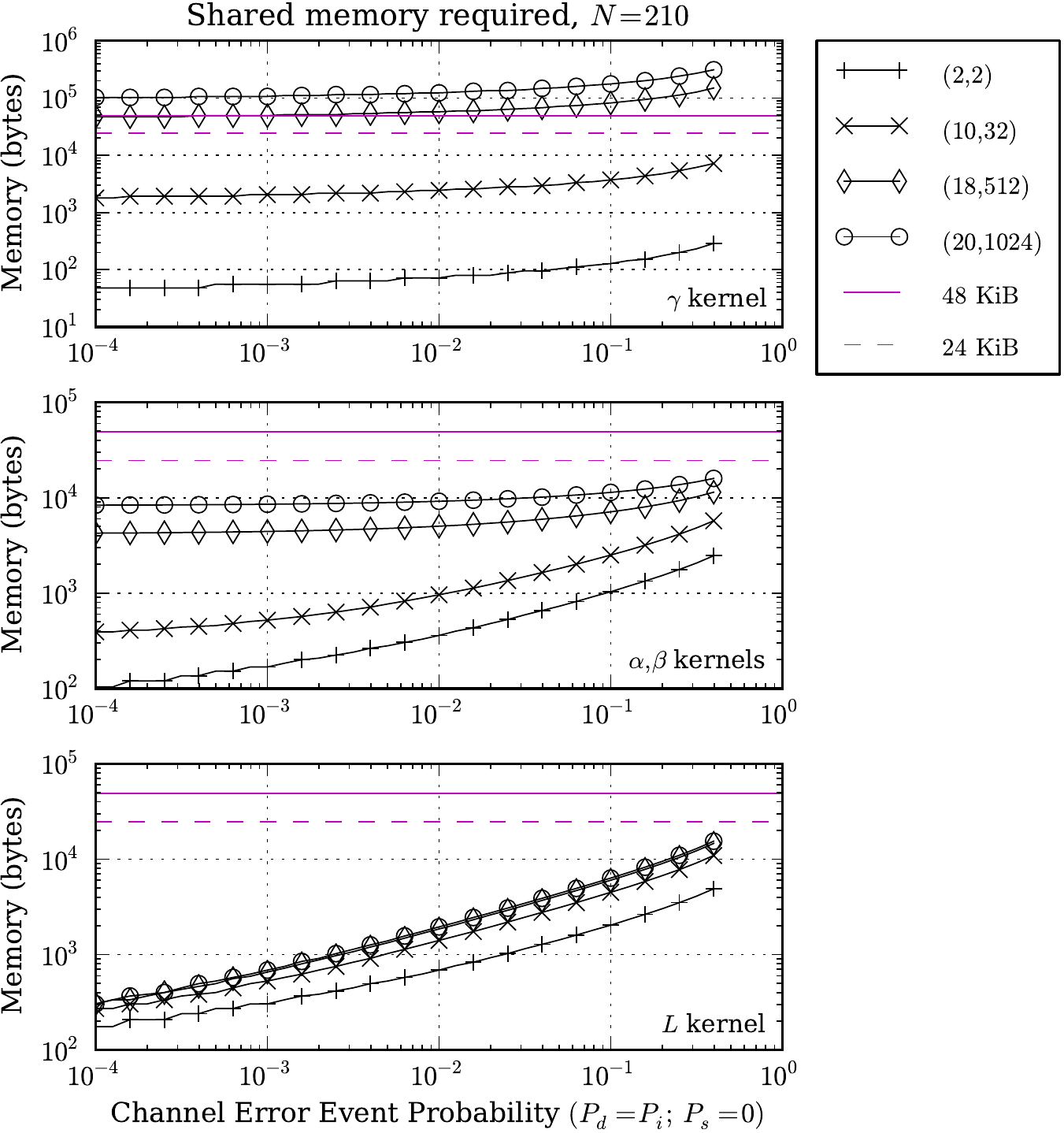}
   {Shared memory requirements per block for the $\gamma$, $\alpha$, $\beta$,
   and $L$ computation kernels, assuming a block size of $q$, for a moderate
   message size $N=210$ and half-rate codewords $(n,q)$, over a range of
   channel conditions.
   It is assumed that internal metrics for the $\gamma$ kernel are stored
   as single precision values, while those for the other kernels are stored as
   double precision values.}
For ease of comparison, horizontal lines are included at values of
$48\,\mathrm{KiB}$ and $24\,\mathrm{KiB}$.
The former corresponds to the maximum amount of shared memory per
multiprocessor on Fermi and Kepler devices; this is the limit for a kernel
launch to succeed.
The latter corresponds to half this value, allowing two resident blocks
per multiprocessor.
Observe how the requirements for the $\gamma$ computation kernel exceed
device limits for larger $q$.
Other kernels only approach device limits when channel conditions are poor;
limits are exceeded if $N$ is sufficiently large.


A kernel launch that violates any of these limits will fail at run-time.
For the four computation kernels, we solve this problem by dynamically
determining a suitable block size at run-time, and adapting the kernel
implementation to work with a block size that is not equal to $q$.
We choose a block size equal to the smaller of $q$ or the largest allowed
multiple of the warp size, taking into account device limits, register
pressure and the shared memory required for a given thread count.
We denote this block size by $B_x^{\gamma}$ for the $\gamma$ computation
kernel, $B_x^{\alpha\beta}$ for the $\alpha$ and $\beta$ computation kernels,
and $B_x^{L}$ for the $L$ computation kernel.
For the $\gamma$, $\alpha$, and $\beta$ computation kernels, if the block
size is less than $q$, each thread loops through values of $D$ equal to its
thread index plus multiples of the block size.
For the $L$ computation kernel we adopt a different approach: for $B_x^{L}<q$,
we extend the grid size by a factor $G_y^{L}=\ceil{q/B_x^{L}}$, dividing the
computation of the range of $q$ across multiple blocks.
This has the advantage of keeping the same parallelization in a kernel which
normally has a small grid size.
%
%
In both cases, unless $q$ is an exact multiple of the block size chosen,
some threads will have no work to do.
In the common case where $q$ is a power of two, any large $q$ is also a
multiple of the warp size (32 for current hardware); in this case there will
be no warp divergence, so any effect on performance should be limited to
the loss in latency hiding.

\subsection{Improving Occupancy}
\label{sec:occupancy}


At the other end of the scale, we had suggested in \cite{briffa13jcomml}
that one may improve performance for small $q$ (and large $N$) by computing
multiple indexes in a single block.
This can potentially improve multiprocessor occupancy, and therefore reduce
latency in memory-limited kernels.
It is worth noting here that occupancy does not depend only on the
block size, but also on the number of resident blocks per multiprocessor
\cite{cuda-bpg-50}.
In turn, this depends on register pressure and shared memory pressure.
Therefore, it is advantageous to increase the block size as long as this
does not increase register and shared memory requirements proportionally.


Consider first the $\gamma$ metric computation.
A suitable strategy for increasing block size is to aggregate the work
done in multiple blocks; rather than aggregating multiple indexes $i$,
however, we aggregate multiple states $m'$.
This has the advantage of allowing the aggregation to happen in both global
and local storage methodologies.
At the limit, one could attempt to use thread coordinates $(D,m')$ for block
size $q \times M_\tau$.
This would require block coordinate $i$ for grid size $N$ in global storage;
there would only be one block in local storage.
However, such a block size is very likely to exceed device limits due to
resource constraints.
Instead, for small $q$, we use a block size $q \times B_y^{\gamma}$, where
$B_y^{\gamma}$ is the smaller of $M_\tau$ or the largest allowed multiple
of the warp size.
In determining $B_y^{\gamma}$, we take into account device limits, register
pressure and the shared memory required for a given thread count.
This results in a grid size $N \times G_y^{\gamma}$ in the case of
global storage or $1 \times G_y^{\gamma}$ for local storage, where
$G_y^{\gamma}=\ceil{M_\tau/B_y^{\gamma}}$.
We also limit $B_y^{\gamma}$ so that the resulting grid size is not smaller
than the number of streaming multiprocessors on the device, $N_\mathrm{SM}$.
Specifically, the constraint is $N G_y^{\gamma} \geq N_\mathrm{SM}$
for global storage, and $G_y^{\gamma} \geq \frac{1}{4} N_\mathrm{SM}$
for local storage (where four such kernels are issued concurrently,
potentially allowing greater occupancy).
This ensures that we do not trade off parallel execution for an increased
occupancy, since the former usually has a greater impact on speed.

A similar argument applies for the $\alpha$ and $\beta$ computation kernels.
For small $q$ we use a block size $q \times B_y^{\alpha\beta}$, where
$B_y^{\alpha\beta}$ is the smaller of $M_\tau$ or the largest allowed multiple
of the warp size.
This results in a grid size
$G_x^{\alpha\beta}=\ceil{M_\tau/B_y^{\alpha\beta}}$.
In addition to considerations for device limits, we limit $B_y^{\alpha\beta}$
so that $G_x^{\alpha\beta} \geq \frac{1}{2} N_\mathrm{SM}$ for global storage
(where $\alpha$ and $\beta$ kernels are computed concurrently), and
$G_x^{\alpha\beta} \geq N_\mathrm{SM}$ for local storage.


We do not apply the same technique to the computation of $L$ since the
proportion of time spent in this kernel is not very significant on the GPU
implementation.
A summary of the kernels required for the complete parallel implementation,
including the advanced considerations described above, for global and local
storage, is given in Table~\ref{tab:kernels-final}.
\inserttabd{tab:kernels-final}{lcccccc}
   {Summary of the kernels required for the complete parallel implementation,
   including considerations for resource limits and multiprocessor occupancy,
   for global and local storage.}{
   \emph{Kernel}
      & \emph{Storage Mode}
      & \emph{Grid size}
      & \emph{Block size}
      & \emph{Calls}
      & \emph{Shared Memory\footnote{%
         Shared memory is expressed here as the unit size of an array
         of real numbers; for the $\gamma$ computation kernel these are
         single precision values, while for all other kernels these are
         double precision values.
         }}
      & \emph{Thread Complexity}
      \\
   \hline
   \multirow{2}{*}{Compute $\gamma$}
      & global
      & $N \times G_y^{\gamma}$
      & $B_x^{\gamma} \times B_y^{\gamma}$
      & $1$
      & \multirow{2}{*}{%
         $(n + m_n^{+} + 1) \cdot B_x^{\gamma} \cdot B_y^{\gamma}$
         }
      & \multirow{2}{*}{%
         $O(n M_n - \frac{m_n^{-} (m_n^{-} - 1)}{2})$
         }
      \\
      & local
      & $1 \times G_y^{\gamma}$
      & $B_x^{\gamma} \times B_y^{\gamma}$
      & $N$
      \\
   \hline
   Initialize $\alpha$,$\beta$
      & both
      & $1$
      & $M_\tau$
      & $1$
      &
      & $O(M_\tau)$
      \\
   Compute $\alpha$,$\beta$
      & both
      & $G_x^{\alpha\beta}$
      & $B_x^{\alpha\beta} \times B_y^{\alpha\beta}$
      & $N$
      & $(q + M_\tau) \cdot B_y^{\alpha\beta}$
      & $O(M_n)$
      \\
   Normalize $\alpha$,$\beta$
      & both
      & $1$
      & $M_\tau$
      & $N+1$
      &
      & $O(M_\tau)$
      \\
   \hline
   \multirow{2}{*}{Compute $L$}
      & global
      & $N \times G_y^{L}$
      & $B_x^{L}$
      & $1$
      & \multirow{2}{*}{$2 M_\tau$}
      & \multirow{2}{*}{$O(M_\tau M_n)$}
      \\
      & local
      & $1 \times G_y^{L}$
      & $B_x^{L}$
      & $N$
      \\
   \hline
   Compute $\Phi_T$
      & both
      & $1$
      & $M_\tau$
      & $2$
      &
      & $O(M_\tau)$
      \\
   }
We also list in the table the complexity of computations for a single thread.


\section{Performance Analysis}
\label{sec:analysis}

We consider the GPU performance of the above implementation on two Fermi
devices and one Kepler device.
The GTX\,480 and GTX\,680 are the highest-performing single-GPU devices in
the consumer-oriented GeForce range for the Fermi and Kepler architectures
respectively.
The C2075 is the highest-performing processor in the computation-oriented
Tesla range for the Fermi architecture.
CPU performance is considered for a serial reference implementation, on the
same system as in \cite{briffa13jcomml}.
Hardware specifications are summarized in Table~\ref{tab:specs}.
\inserttab{tab:specs}{lcccc}
   {Hardware specifications for CPU and GPU systems.}{
      & \emph{Opteron 2431}
      & \emph{GTX\,480}
      & \emph{C2075}
      & \emph{GTX\,680} \\
   \hline
   Processors $\times$ cores 
      & $2 \times 6$
      & $15 \times 32$
      & $14 \times 32$
      & $8 \times 192$ \\
   Core Speed (GHz)
      & 2.412
      & 1.401
      & 1.147
      & 1.0585 \\
   Memory (MiB)
      & 32\,768
      & 1\,536
      & 5\,376
      & 2\,048 \\
   GFLOPs (\texttt{float})
      & 2.412 (scalar) 
      & 672.48
      & 513.856
      & 1\,625.856 \\
   GFLOPs (\texttt{double})
      & 2.412 (scalar) 
      & 336.24
      & 256.928
      & 67.744 \\
   }%

\subsection{Division of Computation Time}
\label{sec:results_division}

We consider first the time spent in each of the four metric computations as
a proportion of the total time required to decode a frame.
For global storage, this is shown in Fig.~\ref{fig:division-global-N-32}
for a moderate alphabet size $q=32$ over a range of message sizes $N$, for
the CPU and the C2075 device.
\insertfig{fig:division-global-N-32}{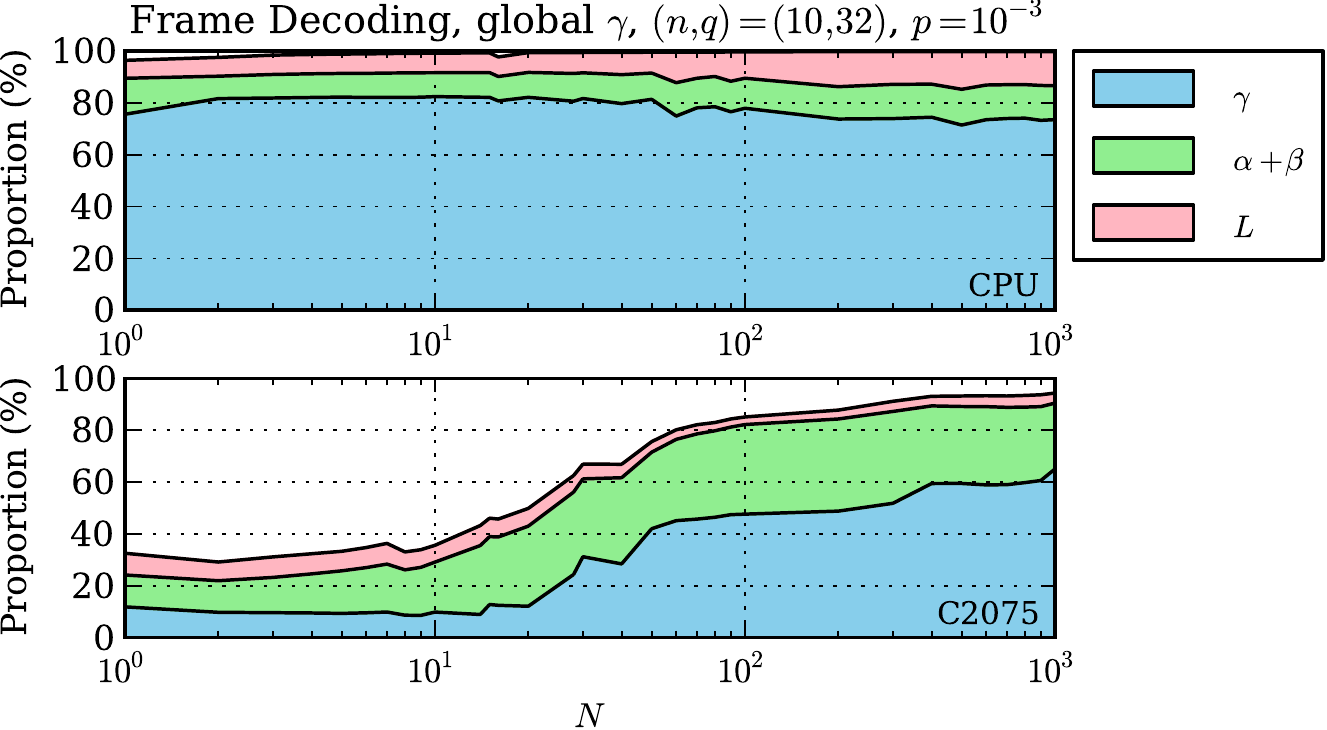}
   {Time spent in each of the four metric computations as a proportion of the
   total time required to decode a frame, for global storage and a moderate
   alphabet size $q=32$, over a range of message sizes $N$.
   Channel conditions are given by $p:=\Pri=\Prd=10^{-3}; \Prs=0$.}
Note that we show a combined timing for the computation of the $\alpha$ and
$\beta$ metrics as these are computed concurrently on the GPU.
On the CPU the division of computation time is almost constant, with the
$\gamma$ metric taking over 75\% of the time for moderate to large $N$.
On the GPU we observe a number of distinct differences:
\begin{inparaenum}[\em a)]
\item the computation of the $\gamma$ metric takes a substantially smaller
proportion of time,
\item the $\alpha$ and $\beta$ metrics seem to make up for most of the
difference for moderate to large $N$, and
\item for smaller $N$ a substantial proportion of time is unaccounted for.
\end{inparaenum}
These differences may be explained as follows.
Computation of $\gamma$ is more easily parallelized than $\alpha$ and $\beta$,
which also suffer from greater kernel call overhead.
This makes the computation of $\gamma$ considerably more efficient than
$\alpha$ and $\beta$ on the GPU.
For smaller block sizes, the overhead in setting up and transferring data to
and from the GPU is also a more significant contributor.
Note that this timing depends on the processor, mainboard, and memory speeds
of the system where the GPU is fitted.
Results for the other two GPU devices show a similar trend.

It is also instructive to repeat this experiment for a moderate message size
$N=210$ over a range of alphabet sizes $q$.
Results for this are shown in Fig.~\ref{fig:division-global-q-210}.
\insertfig{fig:division-global-q-210}{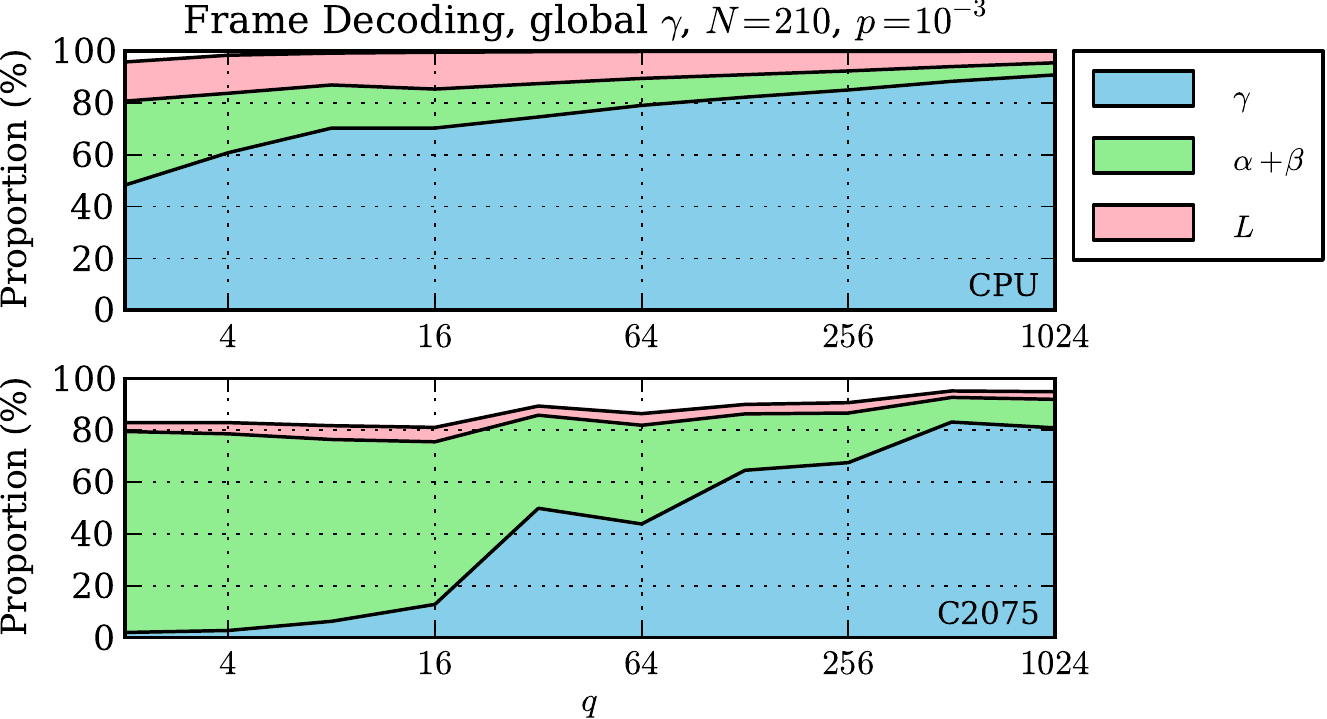}
   {Time spent in each of the four metric computations as a proportion of the
   total time required to decode a frame, for global storage and a moderate
   message size $N=210$, over a range of alphabet sizes $q$.
   Channel conditions are given by $p:=\Pri=\Prd=10^{-3}; \Prs=0$.}
In this case we can see that even on the CPU the time required to compute
$\gamma$ takes a more significant proportion of time as $q$ increases.
The time required by $\alpha$, $\beta$, and $L$ decreases proportionally.
This is expected, as while the computation of $\alpha$, $\beta$, and $L$
scales with $N$, $M_\tau$, $M_n$, and $q$, the computation of $\gamma$
also scales with $n$.
For the codes considered in this experiment, $n = 2 \log_2 q$ to keep the
same code rate.
On the GPU, the change is more pronounced, with the $\gamma$ computation again
becoming dominant for large $q$.
The main reason for this is that as $q$ increases, the computation of $\alpha$
and $\beta$ becomes considerably more efficient, each kernel call has more
computation to perform, and call overhead becomes less significant.
Again, results for the other two GPU devices show a similar trend.

\subsection{Computation Efficiency}
\label{sec:results_efficiency}

We consider next the efficiency of computing the $\gamma$, $\alpha$, and
$\beta$ metrics for the different architectures.
For each metric, this can be visualized by plotting the time taken to compute
the metric, normalized by the expected complexity of computation.
This is shown in Fig.~\ref{fig:normalized-global-N-32} for global storage
and a moderate alphabet size $q=32$ over a range of message sizes $N$.
\insertfig{fig:normalized-global-N-32}{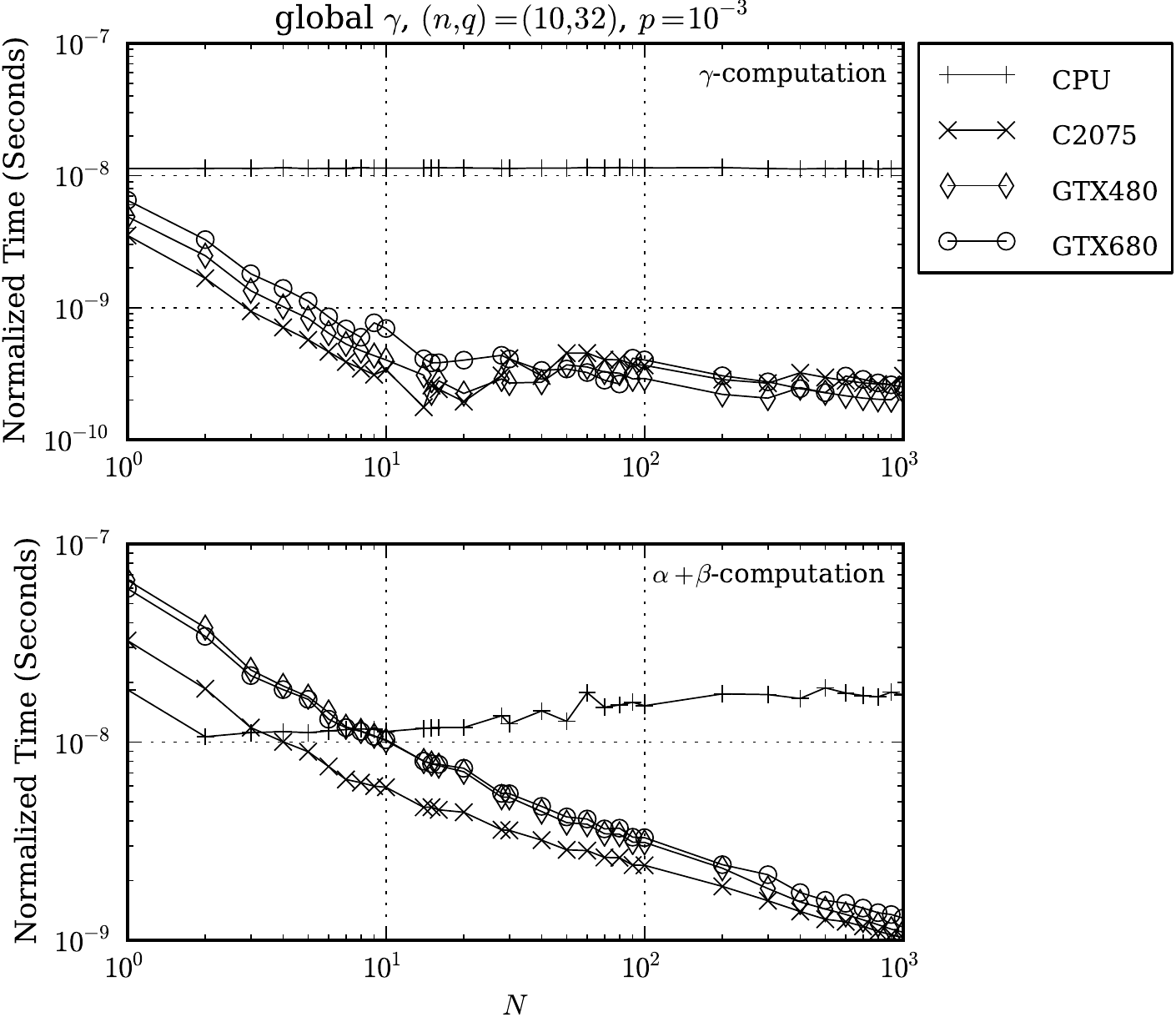}
   {Time to compute
   $\gamma$, normalized by a factor
   $N q M_\tau (n M_n - m_n^{-} (m_n^{-} - 1)/2)$, and
   $\alpha$ and $\beta$, normalized by a factor $N q M_\tau M_n$,
   for global storage and a moderate alphabet size $q=32$, over a range
   of message sizes $N$.
   Channel conditions are given by $p:=\Pri=\Prd=10^{-3}; \Prs=0$.}
Starting with the $\gamma$ metric, observe that the normalized time is
constant for the CPU; this is expected, and indicates that the expected
complexity is an accurate estimate of actual complexity.
This is most likely due to the tight optimization of the implementation.
For the GPU implementation we can see that maximum efficiency is reached
quickly, slighly beyond $N=10$ on all devices.
Efficiency fluctuates beyond this point, though is generally better for
larger $N$.
Comparing with the equivalent plot in \cite[Fig.~1]{briffa13jcomml}, observe
that in this work maximum efficiency is reached much earlier\footnote{%
For the GTX\,480 this was achieved around $N=100$ in \cite{briffa13jcomml}.}.
This is due to the increased grid size (c.f.\ Section~\ref{sec:gamma_global})
and the use of block aggregation to increase occupancy
(c.f.\ Section~\ref{sec:occupancy}).
Note that the normalized time units in this paper and those in
\cite{briffa13jcomml} cannot be compared directly as the complexity is
obtained for different algorithms.

For the $\alpha$ and $\beta$ metrics, we can see in
Fig.~\ref{fig:normalized-global-N-32} that CPU efficiency is almost constant,
decreasing slightly as $N$ increases.
While unexpected, this is not surprising, as the complexity expression
considers only the floating-point arithmetic and ignores overheads of loop
handling and memory access.
On the other hand, GPU efficiency continues to improve as $N$ increases.
This is most likely because an increase in $N$ causes an increase in $M_\tau$
under the same channel conditions; this increases the computation done in
each kernel call, reducing the significance of kernel call overhead.
Furthermore, the efficiency of each kernel call also improves due to the
increased grid size and the use of block aggregation.

The experiment is repeated for a moderate message size $N=210$, over a range
of alphabet sizes $q$.
Results are shown in Fig.~\ref{fig:normalized-global-q-210}.
\insertfig{fig:normalized-global-q-210}{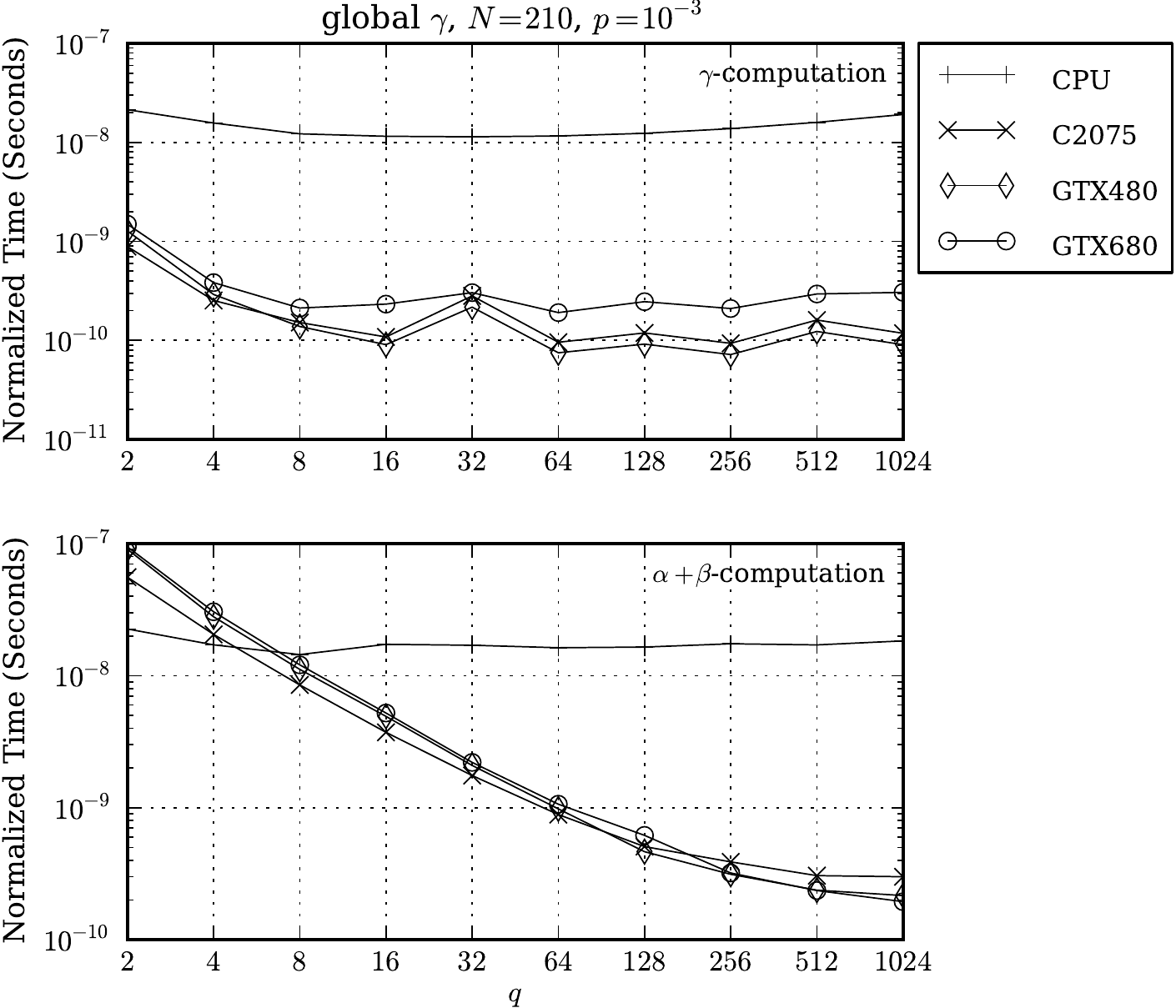}
   {Time to compute
   $\gamma$, normalized by a factor
   $N q M_\tau (n M_n - m_n^{-} (m_n^{-} - 1)/2)$, and
   $\alpha$ and $\beta$, normalized by a factor $N q M_\tau M_n$,
   for global storage and a moderate message size $N=210$, over a range
   of alphabet sizes $q$.
   Channel conditions are given by $p:=\Pri=\Prd=10^{-3}; \Prs=0$.}
In this case, observe that the efficiency of computing the $\gamma$ metric
remains approximately constant throughout the range, with a decrease in
efficiency only for the smallest alphabet sizes.
We can achieve good efficiency throughout the range due to the use of block
aggregation, which increases occupancy and allows good device utilitization
even for small $q$.
%
For the $\alpha$ and $\beta$ metrics, again efficiency continues to improve
as $q$ increases; furthermore, this improvement happens more quickly than
when $N$ is increased (c.f.\ Fig.~\ref{fig:normalized-global-N-32}).
This is because an increase in $q$ has two effects: it directly increases
the block size and indirectly increases the grid size (since $M_\tau$ depends
on $n$, which depends on $q$ if the code rate is fixed).
A similar experiment with larger $N$ shifts the curves to the left, so that
maximum efficiency is reached at a lower $q$.

\subsection{Multiprocessor Occupancy and Usage}
\label{sec:results_usage}

In order to understand the effect of changes in code parameters on computation
efficiency, it is instructive to look at metrics that directly measure the
effectiveness of parallelization.
At multiprocessor level, it is important to keep the hardware busy by having
a sufficiently high thread count per block.
Since instructions are issued at warp level, the block size is ideally a
multiple of the warp size, ensuring that each thread in each warp is doing
something useful.
Furthermore, the hardware hides latencies in instruction issue and memory
access by executing warps that are ready.
It is therefore useful to maximize the number of active warps per
multiprocessor, which can be achieved by increasing the thread count per
block or by ensuring that multiple blocks can be resident.
The latter is often limited by register pressure, so that the former is often
the most practical approach.
Occupancy, the proportion of active warps to the maximum supported by the
hardware, is a useful measure to determine how well latency is hidden.

At device level, effectiveness of parallelization depends on how many of the
available multiprocessors are executing a kernel.
We refer to this proportion as the device usage.
Clearly, if only a single kernel is being executed, usage depends on the
grid size, which is ideally a multiple of the number of multiprocessors
$N_\mathrm{SM}$.
However, this presents a trade-off between the grid size and block size.
Usage can also be improved by running multiple kernels concurrently.
A good strategy is to maximize block size, but only so far as to allow a grid
size which keeps all multiprocessors busy.

For this analysis, we limit our results to the GTX\,480 and GTX\,680 devices.
Multiprocessor occupancy and usage for the C2075 are similar to those of the
GTX\,480, as the two devices have multiprocessors of the same architecture
and only differ in the number of multiprocessors available (by one).

\subsubsection{Global Storage}

Starting first with global storage, we determine occupancy for the $\gamma$,
$\alpha$, $\beta$, and $L$ computation kernels, based on the actual block
size used and assuming no hardware overhead\footnote{%
Note that this is the theoretically achievable occupancy with the chosen
parameters; when measured with the profiler the true value will usually be
marginally lower.
}.
This is shown in Fig.~\ref{fig:occupancy-global-N-32} for a moderate
alphabet size $q=32$ over a range of message sizes $N$.
\insertfig{fig:occupancy-global-N-32}{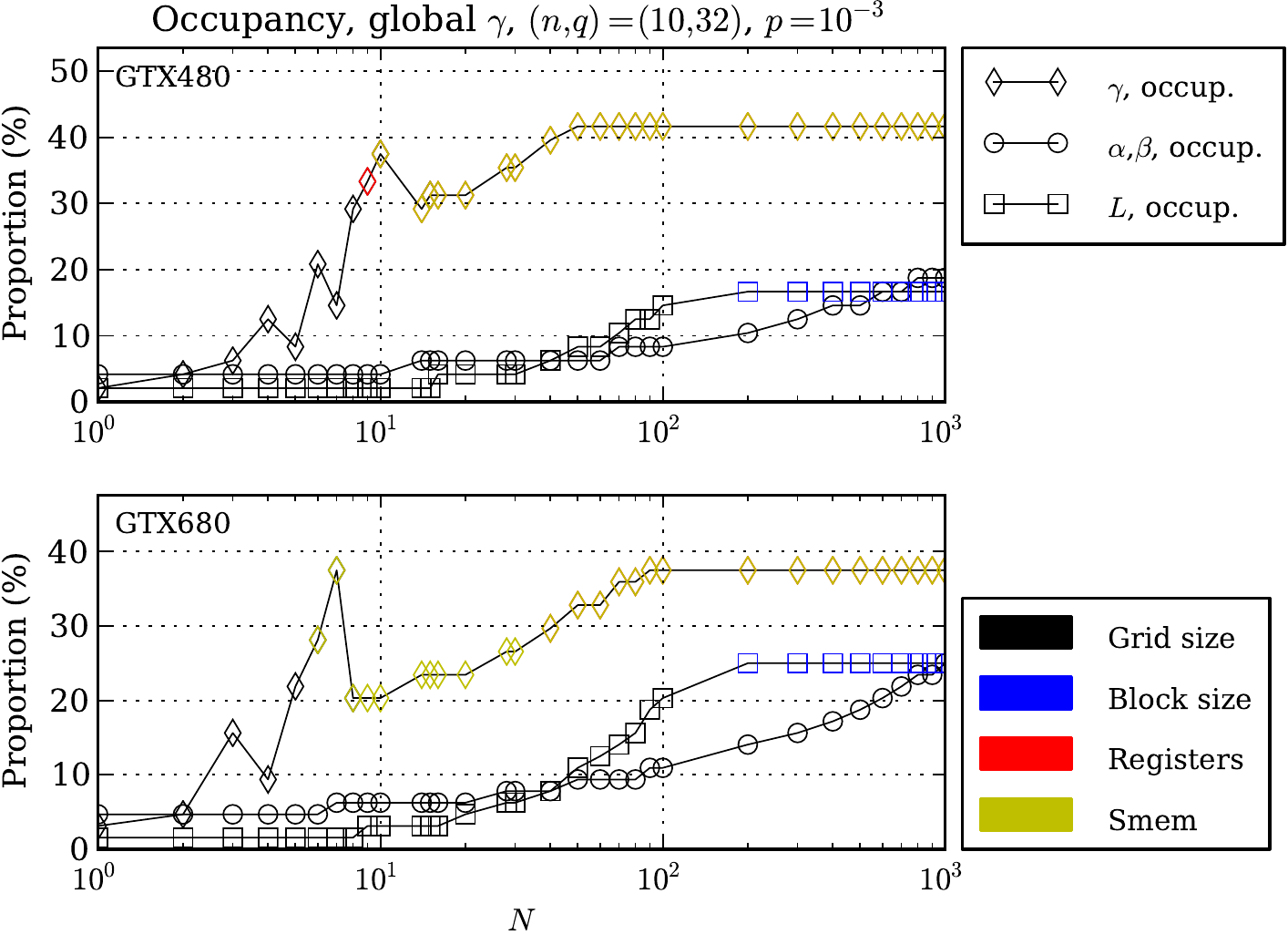}
   {Multiprocessor occupancy for the $\gamma$, $\alpha$, $\beta$, and $L$
   computation kernels for global storage and a moderate alphabet size $q=32$,
   over a range of message sizes $N$.
   Channel conditions are given by $p:=\Pri=\Prd=10^{-3}; \Prs=0$.}
We annotate in color the limiting factor that stops us from achieving higher
occupancy.
For each of the computation kernels, the nominal block size (equal to $q$)
is the same as the warp size.
If this was the chosen block size, occupancy would be limited by the number of
resident blocks.
For Fermi devices this would be at most eight, for a maximum occupancy of 16\%.
Increasing the block size as explained in Section~\ref{sec:occupancy}, however,
allows us to achieve an occupancy of 40\% for the $\gamma$ kernel on Fermi
(and very close to that on Kepler).
Observe that occupancy is limited at the higher end by the shared memory
requirement for the $\gamma$ kernel.
For the $\alpha$ and $\beta$ kernels occupancy is limited by the grid size;
this depends on $M_\tau$ and there is nothing we can do about it.
Comparing these results with Fig.~\ref{fig:normalized-global-N-32}, it
is worth realizing that peak efficiency for $\gamma$ is reached at the same
point when peak occupancy is reached.
Similarly, efficiency for the $\alpha$ and $\beta$ kernels increases with
the kernel occupancy as $N$ is increased.

The corresponding device usage for each kernel is shown in
Fig.~\ref{fig:usage-global-N-32}.
\insertfig{fig:usage-global-N-32}{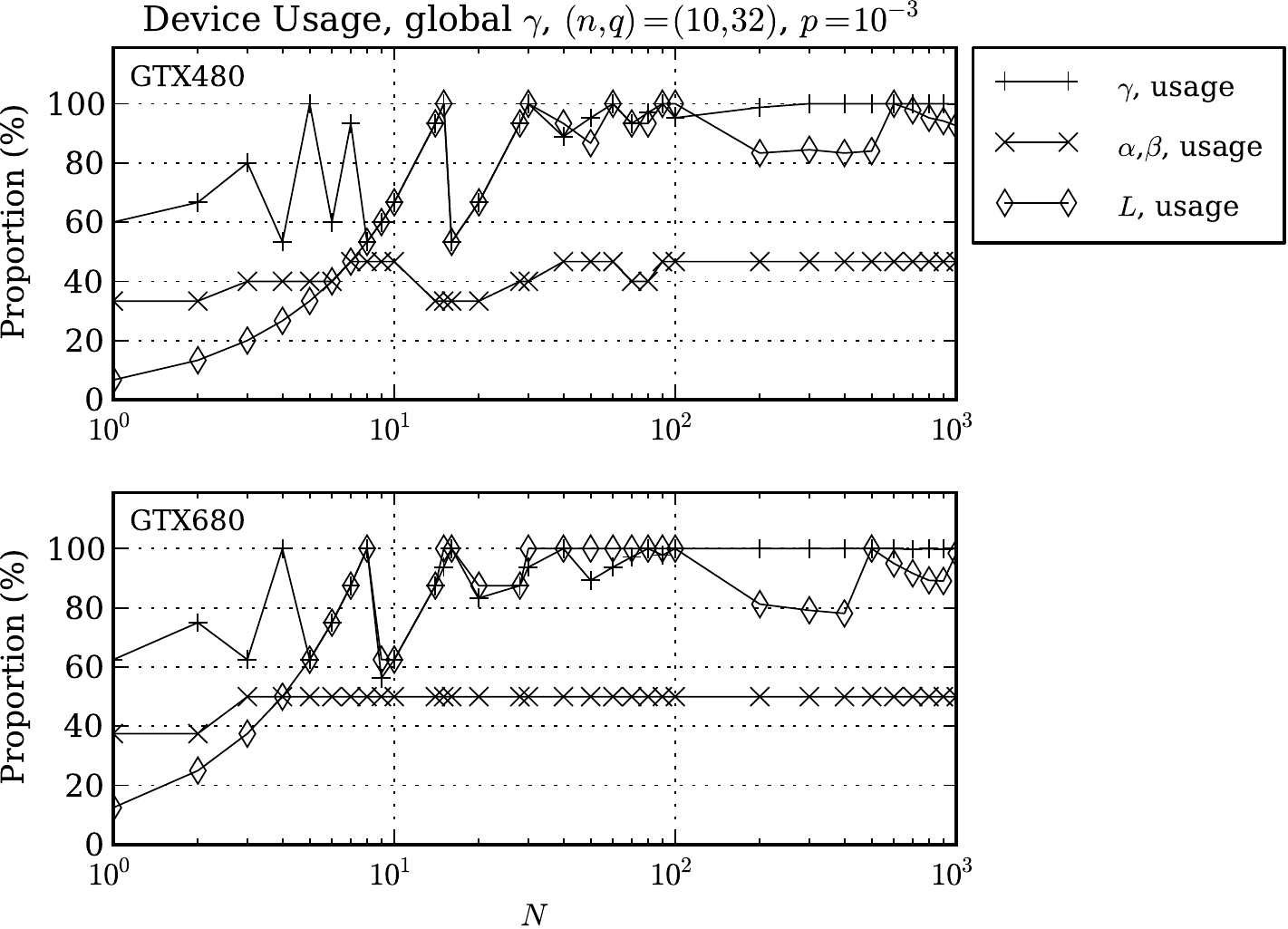}
   {Device usage for the $\gamma$, $\alpha$, $\beta$, and $L$ computation
   kernels for global storage and a moderate alphabet size $q=32$, over a
   range of message sizes $N$.
   Channel conditions are given by $p:=\Pri=\Prd=10^{-3}; \Prs=0$.}
As expected, device usage is optimal for the $\gamma$ and $L$ kernels whenever
$N$ is a multiple of $N_\mathrm{SM}$ for the given device.
Note that for the $\alpha$ and $\beta$ kernels, device usage peaks at 50\% as
we know that these kernels will be running concurrently.
This allows us to reach peak usage earlier for these kernels.
Comparing these results with Fig.~\ref{fig:normalized-global-N-32}, observe
that the fluctuations in device usage for $\gamma$ explain the corresponding
fluctuations in efficiency.

The experiment is repeated for a moderate message size $N=210$, over a range
of alphabet sizes $q$.
Multiprocessor occupancy and device usage are shown respectively in
Fig.~\ref{fig:occupancy-global-q-210}--\ref{fig:usage-global-q-210}.
\insertfig{fig:occupancy-global-q-210}{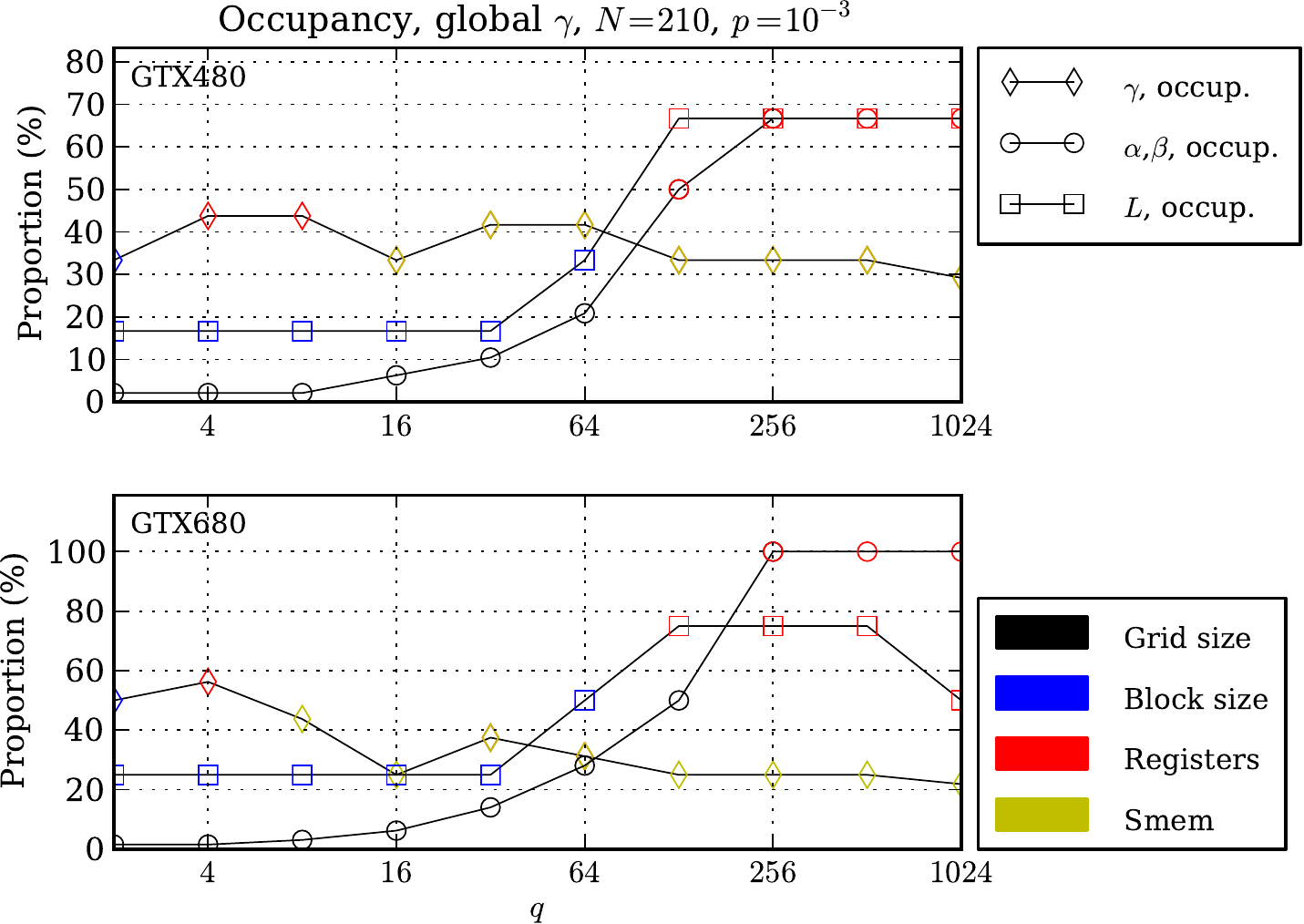}
   {Multiprocessor occupancy for the $\gamma$, $\alpha$, $\beta$, and $L$
   computation kernels for global storage and a moderate message size $N=210$,
   over a range of alphabet sizes $q$.
   Channel conditions are given by $p:=\Pri=\Prd=10^{-3}; \Prs=0$.}
\insertfig{fig:usage-global-q-210}{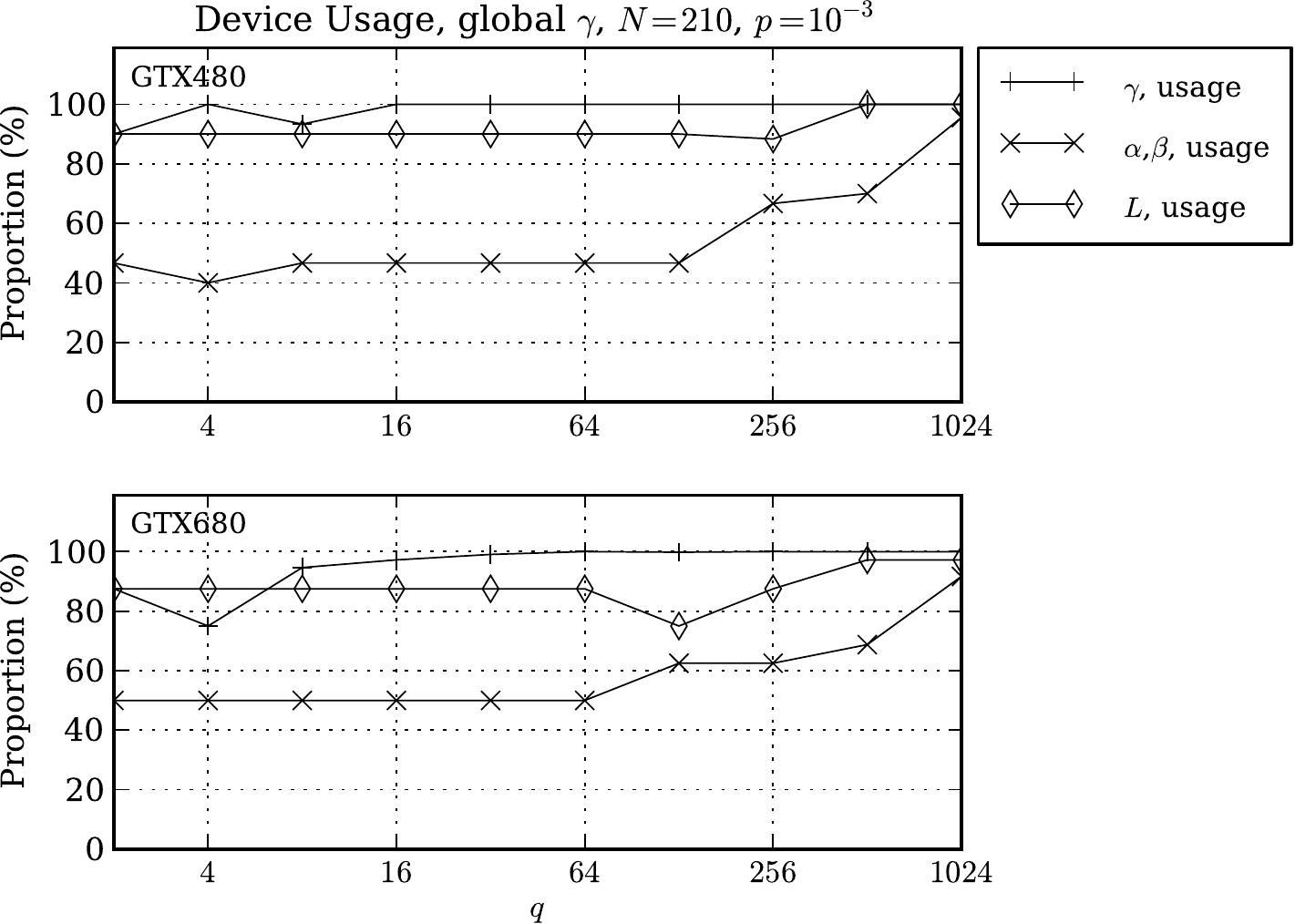}
   {Device usage for the $\gamma$, $\alpha$, $\beta$, and $L$ computation
   kernels for global storage and a moderate message size $N=210$, over a
   range of alphabet sizes $q$.
   Channel conditions are given by $p:=\Pri=\Prd=10^{-3}; \Prs=0$.}
Consider first the Fermi device.
For the $\gamma$ kernel, observe that occupancy remains fairly constant,
limited at the lower end by register pressure, and otherwise by shared
memory requirements.
Together with the constantly high device usage for this kernel, this explains
the flat efficiency in Fig.~\ref{fig:normalized-global-q-210}.
For the $\alpha$ and $\beta$ kernels, however, occupancy increases until it
reaches a peak limited by register pressure.
This is consistent with point where maximum efficiency is reached in
Fig.~\ref{fig:normalized-global-q-210}.

A few points are worth noting with respect to the Kepler device.
This has a higher peak single-precision performance, so one would expect
faster performance for the $\gamma$ kernel.
However, it is much harder to reach the necessary efficiency, for two reasons:
\begin{inparaenum}[\em a)]
\item each multiprocessor has six times as many single-precision cores,
so requires more resident warps to hide latency, and
\item the available shared memory per multiprocessor remains the same,
limiting the occupancy that can be reached.
\end{inparaenum}

\subsubsection{Local Storage}

We repeat the analysis for local storage with a moderate message size $N=210$,
over a range of alphabet sizes $q$.
Multiprocessor occupancy and device usage are shown respectively in
Fig.~\ref{fig:occupancy-local-q-210}--\ref{fig:usage-local-q-210}.
\insertfig{fig:occupancy-local-q-210}{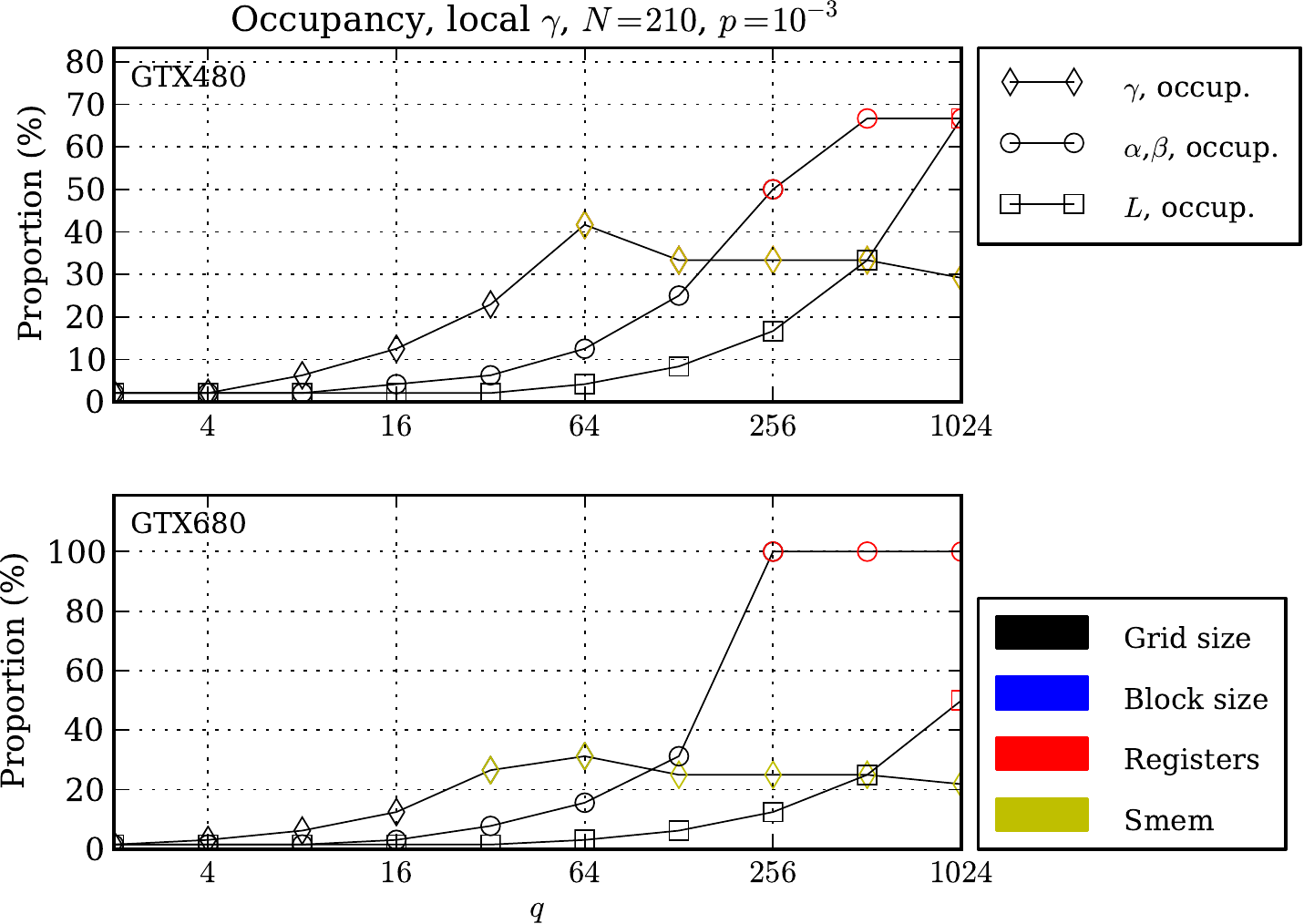}
   {Multiprocessor occupancy for the $\gamma$, $\alpha$, $\beta$, and $L$
   computation kernels for local storage and a moderate message size $N=210$,
   over a range of alphabet sizes $q$.
   Channel conditions are given by $p:=\Pri=\Prd=10^{-3}; \Prs=0$.}
\insertfig{fig:usage-local-q-210}{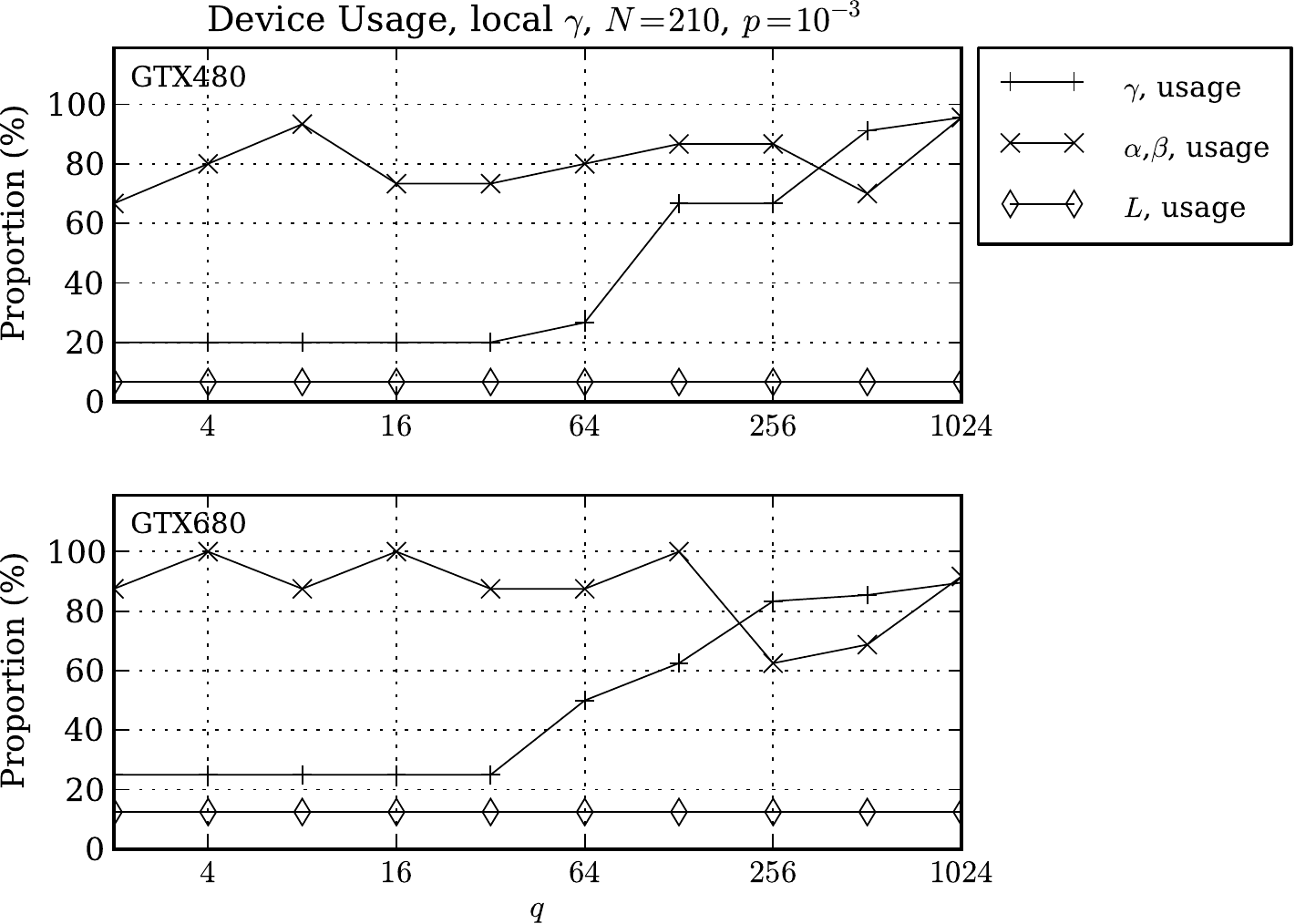}
   {Device usage for the $\gamma$, $\alpha$, $\beta$, and $L$ computation
   kernels for local storage and a moderate message size $N=210$, over a
   range of alphabet sizes $q$.
   Channel conditions are given by $p:=\Pri=\Prd=10^{-3}; \Prs=0$.}
For local storage, each $\gamma$ kernel computes the metric for a single
index $i$; this reduces the grid size by a factor of $N$ in comparison to
global storage.
This places a considerable constraint on the achievable occupancy, especially
if we want to avoid a drop in multiprocessor usage.
For this reason, peak occupancy is only reached for larger $q$.
There is no direct change to the $\alpha$ and $\beta$ kernels with local
storage, except that now we do not run these concurrently.
Therefore, we now try to keep all multiprocessors busy with these kernels
(rather than only half).
This limits the block size by a factor of two, so that we now reach the same
occupancy when $q$ is doubled.

A similar issue occurs with the $L$ metric, which is also computed for a single
index $i$, reducing the grid size by a factor of $N$.
In this case we make no attempt to increase the grid size, relying instead
on the concurrent execution of this kernel with the $\gamma$ and $\beta$
computation kernels.
Given the above observations, we expect the local storage implementation to
be less efficient than global storage, except for large $q$.


\section{Results}
\label{sec:results}

In this section we compare the overall performance of the proposed
implementation on all three GPU systems of Section~\ref{sec:analysis} with
the CPU implementation of the same algorithm and with the earlier GPU
implementation of \cite{briffa13jcomml}.
Following this, we consider the limitations imposed by our implementation
and the overall performance achieved under a range of conditions.

\subsection{Overall Speedup}
\label{sec:results_speedup}

\begin{figure}[tb]
   \centering
   \begin{subfigure}[b]{\figwidth}
      \includegraphics[width=\figwidth]{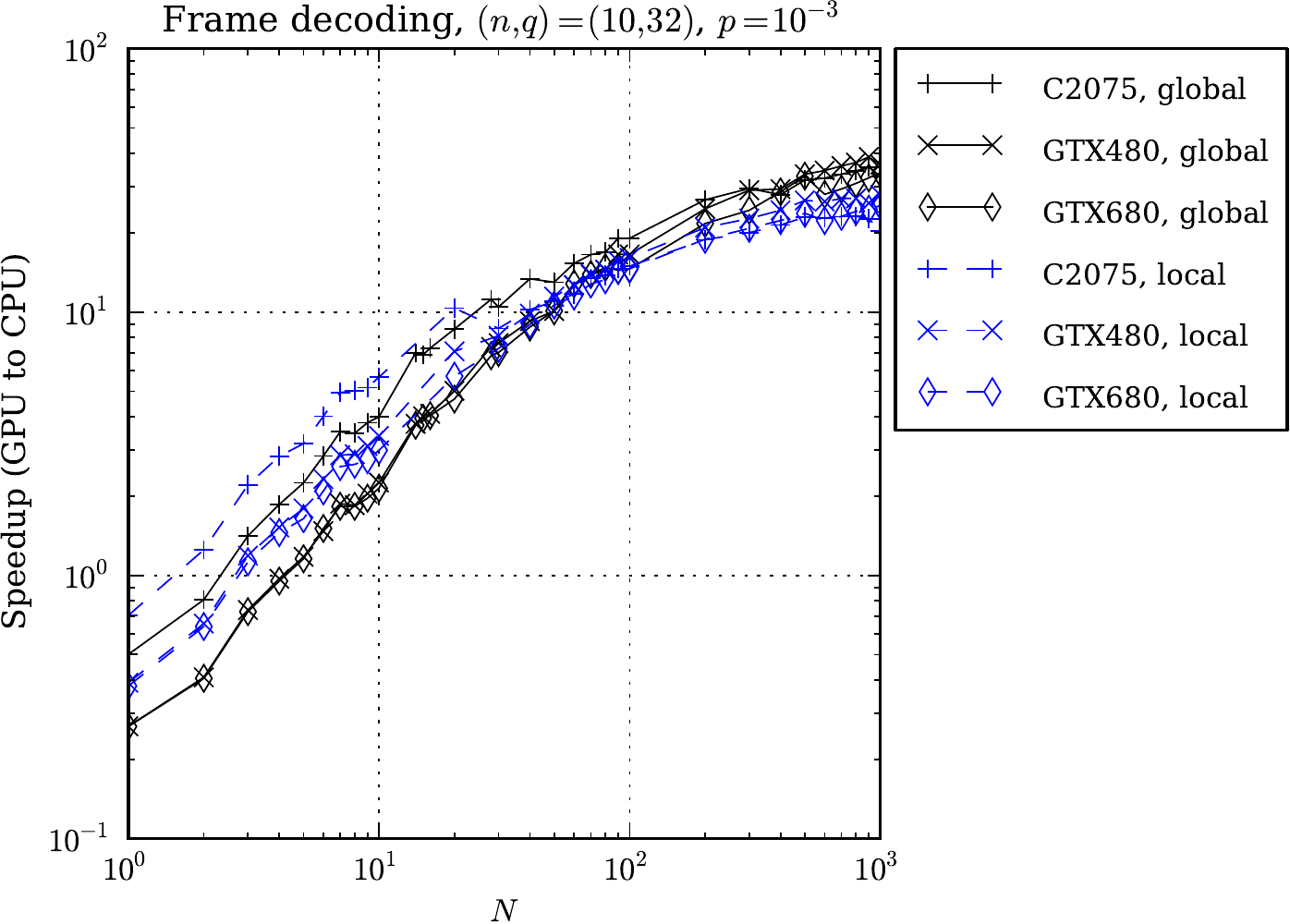}
      \caption{}
      \label{fig:speedup-decoding-N-32}
   \end{subfigure}
   \begin{subfigure}[b]{\figwidth}
      \includegraphics[width=\figwidth]{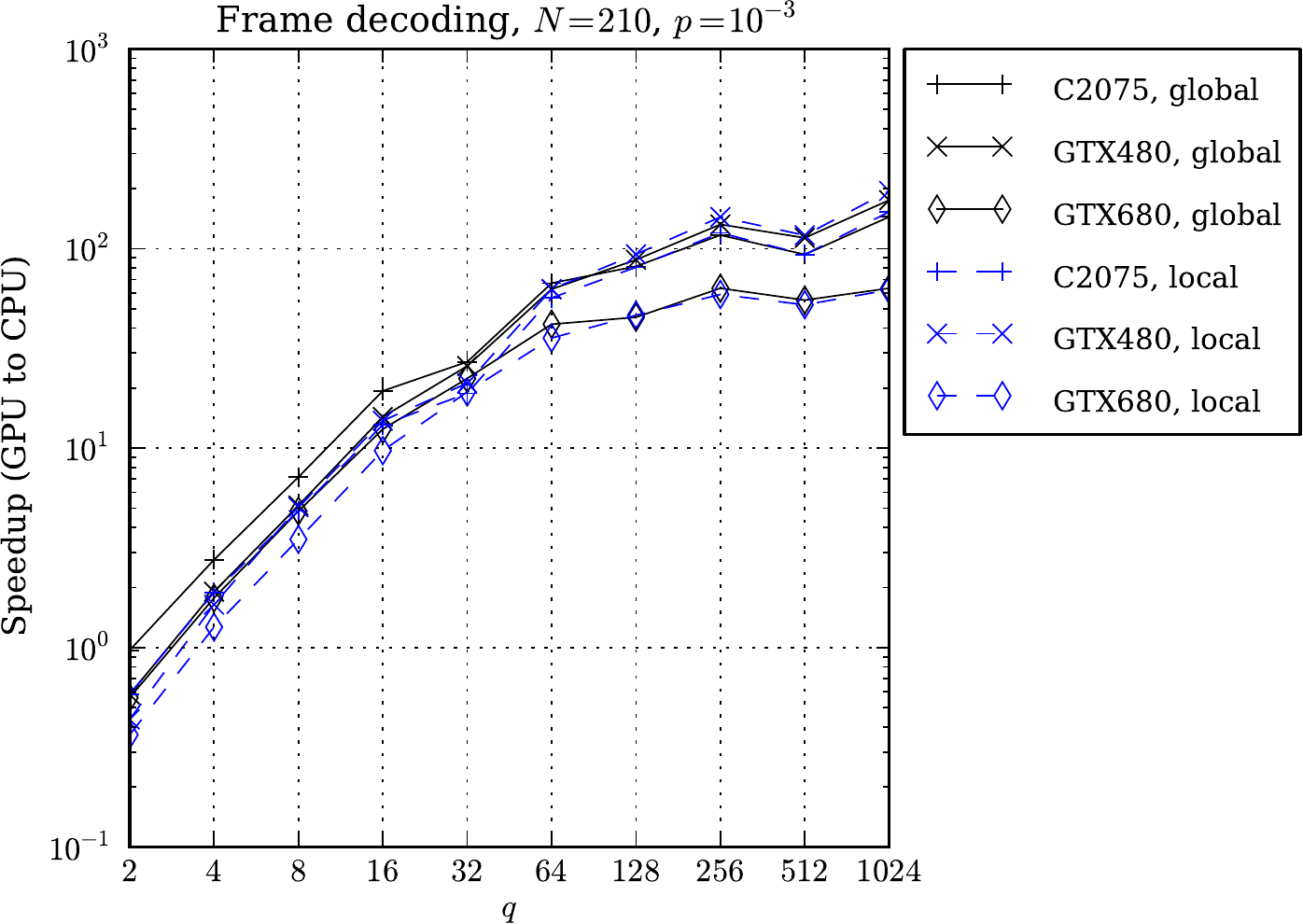}
      \caption{}
      \label{fig:speedup-decoding-q-210}
   \end{subfigure}
   \caption{
      Ratio of complete frame decoding timings on CPU as compared with C2075,
      GTX\,480, and GTX\,680 GPU devices, for global and local storage, with
      \subref{fig:speedup-decoding-N-32}
      a moderate alphabet size $q=32$, over a range of message sizes $N$, and
      \subref{fig:speedup-decoding-q-210}
      a moderate message size $N=210$, over a range of alphabet sizes $q$.
      Channel conditions are given by $p:=\Pri=\Prd=10^{-3}; \Prs=0$.
      }
   \label{fig:speedup-decoding}
\end{figure}

We first consider the speedup for complete frame decoding on the different
GPU devices as compared with the CPU implementation using the same storage
methodology.
This is shown in Fig.~\ref{fig:speedup-decoding} for both global and
local storage.
When we set a moderate alphabet size $q=32$, the speedup improves as the
message size $N$ is increased.
Observe that for small $N$, any improvement is very poor; indeed for the
smallest values of $N$, the CPU implementation is faster than the GPU
implementations.
This is due to the increased impact of GPU overhead when $N$ is small, as
we have seen in Section~\ref{sec:results_division}.
For this value of $q$, observe also that the speedup for global storage is
better than for local storage, for moderate to large $N$.
This is due to the decreased occupancy and multiprocessor usage for small $q$
and low channel error rate, which affects local storage computation in a more
significant way (c.f.\ Section~\ref{sec:results_usage}).

Setting a moderate message size $N=210$, we find that speedup improves as
the alphabet sizes $q$ is increased.
Observe that for small $q$, the speedup achieved for local storage is slightly
less than that for global storage.
For larger $q$, the speedup achieved for local storage is practically the
same as that for global storage.
This indicates the increased efficiency of local storage computation for
larger $q$, as expected.

\subsection{Speedup Over Previous Parallel Implementation}

\begin{figure}[tb]
   \centering
   \begin{subfigure}[b]{\figwidth}
      \includegraphics[width=\figwidth]{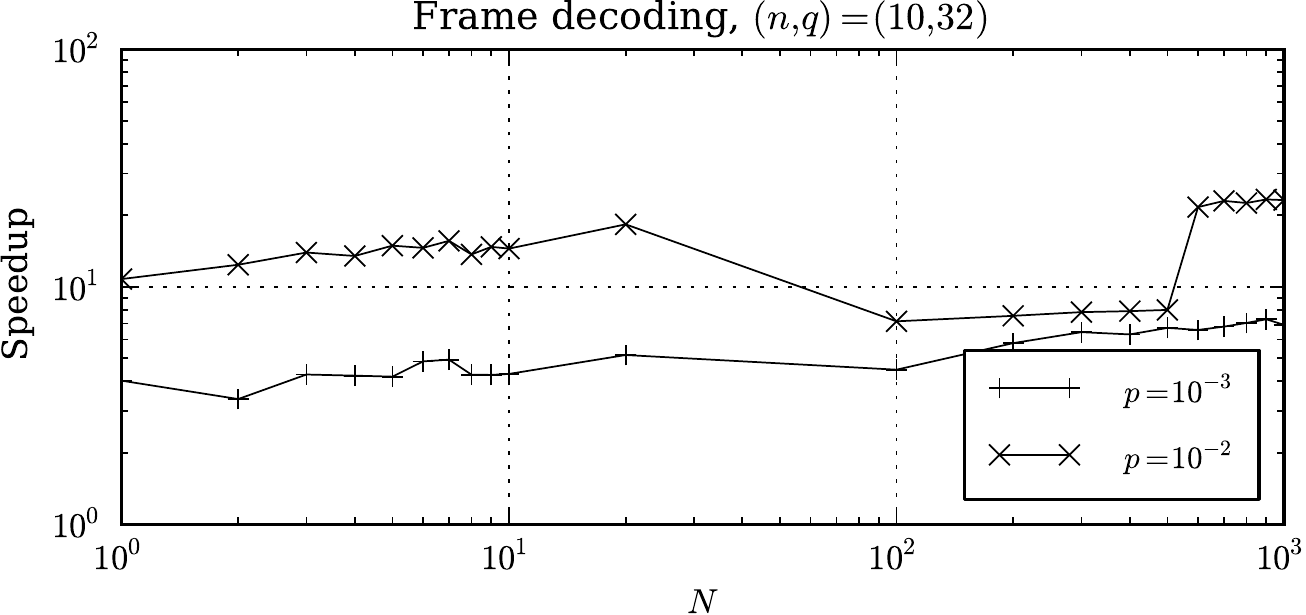}
      \caption{}
      \label{fig:speedup-oldvnew-gpu-N}
   \end{subfigure}
   \begin{subfigure}[b]{\figwidth}
      \includegraphics[width=\figwidth]{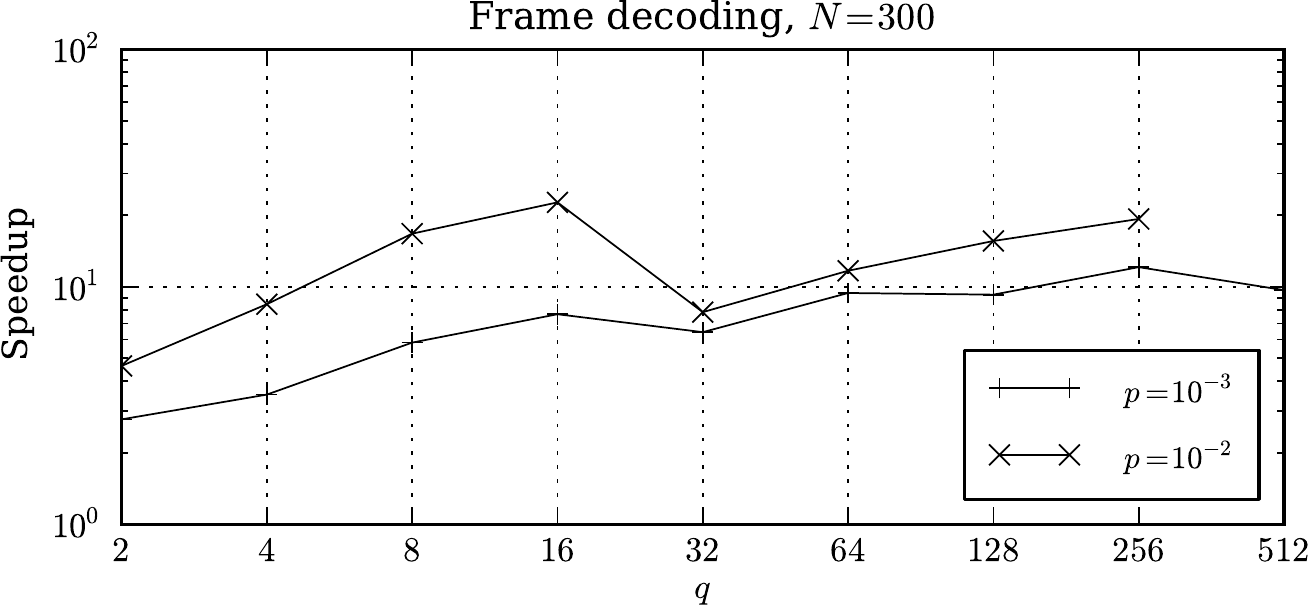}
      \caption{}
      \label{fig:speedup-oldvnew-gpu-q}
   \end{subfigure}
   \caption{
      Decoding speedup for the GPU implementation of this paper (new)
      over the GPU implementation of \cite{briffa13jcomml} (old),
      at different channel conditions for a range of
      \subref{fig:speedup-oldvnew-gpu-N} message size $N$ and
      \subref{fig:speedup-oldvnew-gpu-q} alphabet size $q$.
      Channel conditions are given by $p:=\Pri=\Prd; \Prs=0$.
      }
   \label{fig:speedup-oldvnew-gpu}
\end{figure}

Next, we repeat the simulations of \cite{briffa13jcomml} using the improved
GPU implementation.
Results comparing the overall decoding speed, for the same codes and on the
same GTX\,480 system, are shown in Fig.~\ref{fig:speedup-oldvnew-gpu}.
The speedup obtained for the improved implementation compares favorably
with the expected speedup due to algorithmic changes, shown in
Fig.~\ref{fig:speedup-oldvnew-cpu}.
For larger values of $N$ and $q$ this speedup approaches or exceeds the
speedup achieved on the CPU implementation, indicating that the improved
GPU implementation achieves at least the same efficiency.
This is significant, as a significant fraction of time is spent in computing
the $\alpha$ and $\beta$ metrics, which are much harder to parallelize
efficiently.
As expected, speedup improves for poorer channel conditions.

\subsection{Limitations and Overall Performance}

The main limits imposed by our implementation are discussed below.
The largest supported state space $M_\tau$ is limited by the block size for
the $\alpha$,$\beta$ initialization and normalization kernels and the $\Phi_T$
computation kernel,
Since these kernels are very simple, this limit corresponds to the maximum
number of threads per block, which on current hardware is 1024.
Now $M_\tau$ increases with $\tau$ and the channel error probabilities, and
depends on the allowed probability for an actual drift to exceed these limits.
Even for a poor channel with $\Pri=\Prd=0.2$ and an exclusion probability
$\Prr=10^{-10}$, this allows us to support $\tau \lessapprox 12\,000$.
For a very poor channel having $\Pri=\Prd=0.4$, this limit drops to
$\tau \lessapprox 4\,000$; note that this is beyond the decoding ability of
any known codes.
Larger message lengths are possible when the target channel error rates
are lower.

Another limitation depends on available device memory.
The use of local storage goes a long way to increase the limits of supported
code parameters.
By way of example, a code with $N=840$, $n=20$, $q=1024$ uses a little over
$1\,\mathrm{GiB}$ of device memory at $\Pri=\Prd=0.1$ using local storage.
In general this tends to be an issue only with larger alphabet sizes.

While the implementation considered is intended for use in a simulator, it
is interesting to consider whether the achieved performance is suitable for
real-time applications.
We show in Fig.~\ref{fig:throughput-q-840} the achieved throughput for our
CPU and GPU implementations for global and local storage, a large message
size $N=840$, and a range of alphabet sizes $q$, under moderate to good
channel conditions.
\insertfig{fig:throughput-q-840}{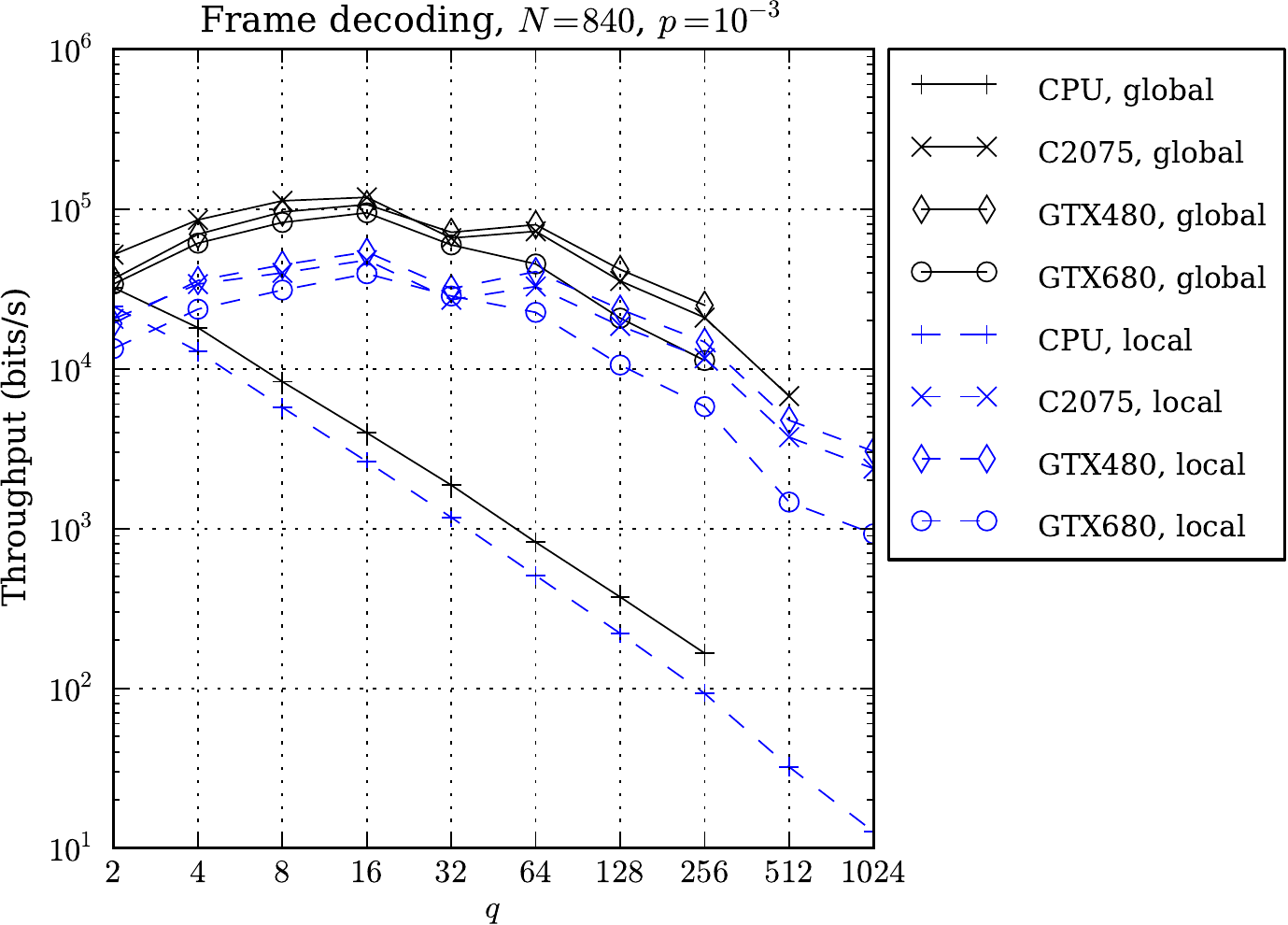}
   {Decoder throughput on CPU and C2075, GTX\,480, and GTX\,680 GPU devices,
   for global and local storage, and a large message size $N=840$, over a
   range of alphabet sizes $q$, under moderate to good channel conditions.
   Channel conditions are given by $p:=\Pri=\Prd=10^{-3}; \Prs=0$.}
Even with the algorithm improvements of \cite{bbw14joe}, the CPU
implementation is clearly too slow for real-time use except in low throughput
applications.
The GPU implementation, however, reaches 100\,kbit/s even with this relatively
large message length.
Note that the missing simulation results for global storage with large $q$
indicate conditions where available memory was exceed.


\section{Conclusions}
\label{sec:closure}

In this paper we have presented an optimized parallel implementation of the
MAP decoder of \cite{bbw14joe} with algorithmic improvements over the
equivalent decoder of \cite{bsw10icc}.
This implementation achieves a speedup of more than $100\times$ over the
CPU implementation of the same algorithm, and more than $10\times$ over the
previous GPU implementation of \cite{briffa13jcomml}, on the same hardware.
This increases our ability to analyse larger codes and poorer channel
conditions, and makes practical use of such codes more feasible.

We also present a reduced-memory implementation where some intermediate
metrics are computed twice: once for the forward pass and once again for
the backward pass and final results.
This variant trades off some decoding speed for significantly reduced memory
requirements.
This allows the decoder to work with longer message sizes and poorer channel
condition than would otherwise be possible.

The speed improvements of this implementation are made possible by a number
of specific optimizations.
We use shared memory judiciously to reduce global memory transfers and to
improve memory access patterns.
We also introduce a dynamic strategy for choosing kernel block sizes,
ensuring efficient use of device resources under a wide range of code and
channel parameters.
Specifically, we determine settings that maximize occupancy while avoiding
idle time on multiprocessors.
Taking these decisions at run-time also has the advantage that this
automatically caters for different devices.
We hope that other researchers working with CUDA will also find these
techniques relevant to their work.

\balance

While this implementation represents a considerable improvement on earlier
implementations, some aspects bear further analysis.
In particular, efficiency of the parallel implementation for small alphabets
is still significantly lower than for larger alphabets.
Aggregation of work previously done on multiple blocks into a single block
has helped by improving occupancy.
However, not all kernels can be improved equally, and for smaller alphabets
a greater proportion of decoding time is spent in the $\alpha$ and $\beta$
metric computations.
This effect is more pronounced under better channel conditions, when the
required state space is much smaller.
Another aspect that may be improved is the optimization of multiprocessor
usage.
While our dynamic strategy for choosing kernel block size ensures that
the usage is never low, we do not seek to optimize the value, prioritizing
instead a higher occupancy.
Of course, improving usage is further complicated by the concurrent execution
of kernels of different size and duration.
However, it may be possible to increase overall efficiency by adapting our
dynamic strategy to take into account empirical kernel timings, which may,
for example, be obtained during system initialization.
Clearly, neither of the above is a trivial proposition; both are the subject
of further work.

\section*{Acknowledgment}
The author would like to thank Prof.~Ing.~V.~Buttigieg and Dr~S.~Wesemeyer
for helpful discussions and comments.


\bibliographystyle{IEEEtran}
\bibliography{IEEEabrv,jbabrv,turbo,unsorted,victor,jb,jbworking}

\end{document}